\documentclass[letterpaper,10pt,twoside]{article}
\usepackage{appendix}

\usepackage{hyperref}
\hypersetup{
	colorlinks,
	citecolor=blue,
	filecolor=blue,
	linkcolor=blue,
	urlcolor=blue,
	linktocpage=true
}

\usepackage{amsmath}
\usepackage{mathptmx}
\usepackage{amsthm}
\usepackage{amssymb}
\usepackage{amsfonts}
\DeclareMathAlphabet{\mathcal}{OMS}{cmsy}{m}{n}

\usepackage[top=1.1in,left=1in,right=1in,bottom=1.1in]{geometry}

\usepackage{graphicx}

\usepackage[rgb,dvipsnames]{xcolor}

\newcommand\mc[1]{\mathcal{#1}}
\newcommand{\tr}{\text{tr}}
\newcommand{\Tr}{\text{Tr}}

\newcommand{\gra}{\alpha}

\newcommand{\grb}{\beta}

\newcommand{\grd}{\delta}

\newcommand{\grf}{\phi}
\newcommand{\grF}{\Phi}

\newcommand{\grl}{\lambda}

\newcommand{\grS}{\Sigma}

\usepackage{filecontents}
\usepackage{breqn}
\usepackage{cancel}
\usepackage{mathtools} 
\usepackage{extarrows}
\usepackage{subfig}

\usepackage[utf8]{inputenc}

\usepackage[compress,numbers]{natbib}

\newcommand{\beq}{\begin{equation}}
\newcommand{\eeq}[1]{\label{#1}\end{equation}}
\newcommand{\ber}{\begin{eqnarray}}
\newcommand{\eer}[1]{\label{#1}\end{eqnarray}}

\newcommand{\al}{\alpha}
\newcommand{\be}{\beta}
\newcommand{\ga}{\gamma}
\newcommand{\G}{\Gamma}
\newcommand{\de}{\delta}
\newcommand{\ta}{\theta}

\newcommand{\pa}{\partial}
\newcommand{\na}{\nabla}
\newcommand{\nabp}[1]{\nabla'{}_{\!\!#1}}
\newcommand{\ab}{{\alpha\beta}}

\newcommand{\then}{~~~\Rightarrow~~~}
\usepackage{bbm}
\newcommand\id{\ensuremath{\mathbbm{1}}} 

\usepackage{amsbsy}
\renewcommand{\epsilon}{\text{\usefont{OML}{cmr}{m}{n}\symbol{15}}}

\begin{document}
\renewcommand{\theequation}{\thesection.\arabic{equation}}
\setcounter{page}{0}

\thispagestyle{empty}

\begin{flushright} \small
YITP-SB-18-20\\
\end{flushright}
\vskip15mm
\begin{center}
{\huge\bf Dualities and Phases of $3d$ $\mc{N}=1$ SQCD} \\
\vskip10mm
{\Large Changha Choi${}^1$, Martin Ro\v{c}ek${}^{2}$, 
Adar Sharon${}^3$} \\
\vskip15mm
{${}^1$Physics and Astronomy Department, Stony Brook University, Stony Brook, NY 11794 USA} \\
\vskip5mm
{${}^2$C.~N.~Yang Institute for Theoretical Physics, Stony Brook University \\ Stony Brook, NY 11794 USA} \\
\vskip5mm
{${}^3$Department of Particle Physics and Astrophysics, Weizmann Institute of Science, Rehovot, Israel}
\vskip10mm
\end{center}

\date{\today}
\vskip1cm
\begin{center}
{\large\bf Abstract} \\[5mm]
\end{center}
We study gauge theories with $\mc{N}=1$ supersymmetry in 2+1 dimensions. We start by calculating the 1-loop effective superpotential for matter in an arbitrary representation. We then restrict ourselves to gauge theories with fundamental matter. Using the 1-loop superpotential, we find a universal form for the phase diagrams of many such gauge theories, which is proven to persist to all orders in perturbation theory using a symmetry argument. This allows us to conjecture new dualities for $\mc{N}=1$ gauge theories with fundamental matter. We also show that these dualities are related to results in $\mc{N}=2$ supersymmetric gauge theories, which provides further evidence for them.
\vskip6cm
{\small Email: \texttt{Changha.Choi@stonybrook.edu,
Martin.Rocek@stonybrook.edu, adarsharon1@gmail.com}}
\newpage
\setcounter{tocdepth}{1}
\tableofcontents
\appendixtitleon
\appendixtitletocon
\newpage
\setcounter{tocdepth}{2}

\section{Introduction and Summary}
\setcounter{equation}{0}

Three-dimensional gauge theories have been the focus of many recent studies. A wide array of tools has been applied to these theories, leading to many exact results and even more conjectures. However, the $\mathcal{N}=1$ supersymmetric\footnote{2+1$d$ Theories with $\mc{N}=1$ supersymmetry have 2 supercharges.} versions of these theories have only very recently received increased attention \cite{Bashmakov:2018wts,Benini:2018umh,Gomis:2017ixy,Gaiotto:2018yjh,Eckhard:2018raj,Inbasekar:2015tsa,Benini:2018bhk}. These theories are the focus of the present paper.

Supersymmetry (SUSY) is less restrictive in 2+1$d$ $\mc{N}=1$ theories than in other well-known SUSY theories. For example, since the superpotential is real, the standard non-renormalization theorems based on holomorphy, which apply to many SUSY theories (such as $\mc{N}=1$ SUSY in 3+1$d$), do not apply to $\mc{N}=1$ theories in 2+1$d$. These theories are therefore much harder to analyze than theories with more supercharges.
However, many properties of SUSY do not depend on holomorphy of the superpotential, and they can come to our aid when studying $\mc{N}=1$ theories in 2+1$d$. The Witten index \cite{Witten:1982df} is one example; we rely on it heavily. Phase transitions between SUSY vacua are also simpler to analyze, since these phase transitions must be of second order or higher\footnote{This is because SUSY vacua have zero energy. Recall that classical SUSY vacua with vanishing Witten index can be lifted dynamically, whereas classical SUSY vacua with non-vanishing Witten index must have zero energy also at the quantum level.}. One final example is the calculation of supergraphs, which makes cancellations between bosonic and fermionic Feynman diagrams manifest, and which simplify our calculations. These tools allow us to find some exact results even though the superpotential is real.

In this work we use supergraphs to find the phase diagrams of $\mathcal{N}=1$ gauge theories coupled to fundamental matter multiplets. When the theories are weakly coupled (e.g. when the fields are very massive), the vacuum can usually be found through a simple semiclassical computation. However, in the strongly coupled regime the vacua are more complicated. In this case one must calculate the effective potential of the theory to find the correct vacua \cite{PhysRevD.7.1888}, which can be calculated in perturbation theory using Feynman diagrams. We use a SUSY generalization of this effective potential (called the effective superpotential).

Given a 2+1$d$ $ \mc{N}=1 $ gauge theory with massive matter fields $\grF$, we calculate the 1-loop effective superpotential for matter fields in an arbitrary representation $R$, assuming that there is a single Chern-Simons (CS) level\footnote{By this we mean that the result is valid when the "quantum" CS levels (schematically obtained by integrating out all of the fermions, see e.g. \cite{Komargodski:2017dmc}) are all equal. This is automatically obeyed for gauge groups such as $SU(N),O(N),SO(N),Sp(N)$ and so on, but is also obeyed for theories such as $\mc{N}=1$ $U(N)_{k+\frac12 N,k}$ (since after integrating out the gaugino we find $U(N)_k$, which has a single CS level).} $k$. The result is\footnote{Our convention here is that under time reversal, $k\rightarrow -k$, so it suffices to study $k>0$.}
\begin{equation}\label{eq:intro_superpotential}
W= m|\grF_i|^2-\frac{\kappa}{8\pi}\tr \sqrt{\kappa^2\delta^{ab}+4g^2\bar\Phi_i T^{(a} T^{b)} \Phi_i}
\end{equation} 
Here $m$ is the tree-level mass, $\kappa=\frac{kg^2}{2\pi}$, $g$ is the gauge coupling, $i=1,...,N_f$ is the flavor index and $T^a$ for $a=1,..,\dim(R)$ are the generators of the gauge group in the representation $R$.

Next, we restrict ourselves to matter in the fundamental representation; we mostly discuss the unitary, orthogonal and symplectic groups in this paper. Having found the effective superpotential, we can study the vacua and the phase structure of the theory. We find that 1-loop corrections to the superpotential have a drastic effect on the phase diagram of the theory. While the classical theory can have a moduli space (when the matter fields are massless), this moduli space is generally lifted at the 1-loop level. On the other hand, the 1-loop superpotential can create new vacua that do not appear at the classical level when the fields are massive.

\begin{figure}[ht]
\centering
\includegraphics[width=0.7\linewidth]{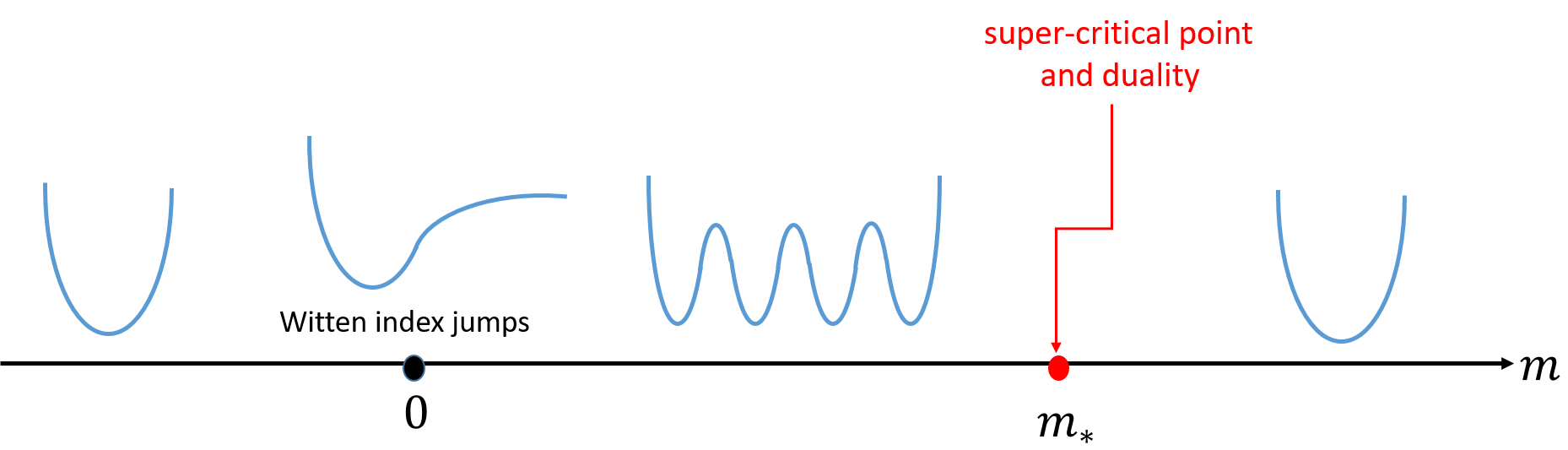}
\caption{\small Schematic general form of the phase diagram for the $\mathcal{N}=1$ gauge theories with CS terms and $N_f$ fundamental matter fields discussed in this paper. The blue graphs depict the effective potential. There are two semiclassical phases for large $|m|$, and an intermediate phase with $\min(N+1,N_f+1)$ solutions to the F-term equations with various symmetry breaking patterns. A similar phase diagram exists for $SO(N)$ and $Sp(N)$ gauge groups. The phase transition at $m=m_*$ is highly unnatural, since at this point many solutions of the F-term equations collide simultaneously. Nonetheless, this picture is an exact result to all orders in perturbation theory.}
\label{fig:general_phase_diagram}
\end{figure}

We summarize our results for the phase diagrams of these theories. We find that the theories we study all have a universal form for their phase diagram shown in Figure \ref{fig:general_phase_diagram}.
For concreteness, we describe the phases for a specific example we study in this paper: a $U(N)_{k+\frac12(N+N_f),k+\frac12 N_f}$ gauge theory with $ N_f $ fundamental matter multiplets. Solving the F-term equations, we find the phase diagram that appears in Figure \ref{fig:general_phase_diagram}, with the following phases:
\begin{itemize}
    \item For $m>m_*$ for some $m_*>0$, we find a single solution to the F-term equations. The resulting low-energy theory is an $\mc{N}=1$ $U(N)_{k+\frac{N}{2}+N_f}$ CS theory.
    \item For $m<0$, we again find a single solution, with an $\mc{N}=1$ $U(N)_{k+\frac{N}{2}}$ CS theory.
    \item In the intermediate phase $0<m<m_*$, we find $\min(N+1,N_f+1)$ solutions to the F-term equations. These exhibit various symmetry-breaking patterns, and consist of an $\mc{N}=1$ vector multiplet with a decoupled non-linear sigma model (NLSM).
\end{itemize}
We claim that this picture is correct for all $N$ and $N_f$ in the range $k>-\min(N_f,N)$. Note that some of the vacua above may have vanishing Witten index, and are expected to break SUSY dynamically due to non-perturbative corrections. This behavior of a pure $\mc{N}=1$ CS theory is reviewed in Section \ref{free_N1_vector_multiplets}.

There are two interesting points in the phase diagram that we must discuss. First, at the point $m=0$ we find a "wall" where the Witten index jumps due to the appearance of vacua from infinity \cite{Witten:1982df} (this appeared in a similar analysis where the matter content was a single adjoint matter multiplet \cite{Bashmakov:2018wts}). Second, we find a single phase transition point where many SUSY vacua merge together to create a single SUSY vacuum (specifically, for $k\geq 0$ we find that the number of SUSY vacua that merge is $\min(N_f+1,N+1)$). This critical point is highly unnatural; generically, we would need to tune $n-1$ relevant deformations to find a fixed point where $n$ vacua collide. Note that such a critical point did not appear when the matter field was in the adjoint representation; there, instead of one non-generic critial point, there were many different phase transitions in the range $0<m<m_*$ for some $m_*$ \cite{Bashmakov:2018wts}. We  call the non-generic critical point at $m=m_*$ a "super-critical" point\footnote{This term may lead one to suspect that SUSY is required for this phenomenon to occur. We are not claiming that this is true in the general case. However, we find that for the theories we study here, there is strong evidence that the super-critical points can be linked to similar phenomena in 3$d$ theories with $\mc{N}=2$ SUSY (where they can be understood as resulting from the holomorphy of the superpotential).}.

The phase diagram can be explicitly found at 1-loop by studying the effective superpotential \eqref{eq:intro_superpotential}. However, we show that this phase diagram is actually an exact result to all orders in perturbation theory (and only the value of $m_*$ can be changed by higher-loop corrections). This makes the phase transition point at $m_*$ much more interesting. As explained above, having such a super-critical point in the theory is highly unnatural, since a generic small quantum correction should split this single point into many phase transitions. However, we are able to use general arguments that rely only on the symmetries of the theory to show that this super-critical point should indeed persist to all orders in perturbation theory.

These symmetry arguments turn out to useful in an even wider range of theories.
Much of this paper is devoted to studying the solutions of the F-term equations due to the superpotential \eqref{eq:intro_superpotential} in the range $0<m<m_*$ for various gauge groups. As discussed above, we find that the solutions are universal, which means that a wide range of gauge theories with CS terms and fundamental matter fields have the same general form for their phase diagrams, appearing in Figure \ref{fig:general_phase_diagram}.
We explain this universality using the same symmetry arguments, and show that these solutions are expected to persist to all orders in perturbation theory, again allowing for exact results for the theories we study in this paper.

Next, we study the phase transition point at $m_*$ in more detail. As explained above, since the phase transition is between SUSY vacua, it must be (at least) second order. The phase transition is thus described by a conformal field theory (CFT). We can study some of its properties by considering the RG flows of these theories at large $k$. The two-loop beta functions of CS theories with matter were calculated in \cite{AVDEEV1992561,avdeev1993renormalizations}. However, since our theories also include a Yang-Mills (YM) term, it is not immediately obvious how to relate these results to our theory. To do so, we must first integrate out the very massive YM mode from our theory, which lead to a low energy CS-matter theory with some effective superpotential. We can then use the two-loop beta functions of \cite{AVDEEV1992561,avdeev1993renormalizations} to find the fixed point we flow to, and to study the corresponding CFT at the super-critical point. In particular, it is shown that while this fixed point generically has $\mc{N}=1$ SUSY, in some degenerate cases we find emergent $\mc{N}=2$ SUSY in the infrared (IR). The special case of an $SU(2)$ gauge theory with $N_f$ fundamental matter multiplets is discussed separately. The global symmetry of this theory is classically enhanced\footnote{To avoid confusion, we use the word "enhanced" when discussing symmetries which are classically larger than naively expected, and "emergent" when discussing symmetries which appear only in the IR. Specifically, while an $SU(N)$ gauge theory with $N_f$ fundamental matter fields has a $U(N_f)$ global symmetry, in the case $N=2$ this is classically enhanced to $Sp(N_f)$, as we shall review in this paper.}  to $Sp(N_f)$, and we find that this enhanced symmetry is present in the IR fixed point as well.

The phase diagrams and RG flows lead us to propose some $\mc{N}=1$ dualities at the phase transition point $m=m_*$. A duality was recently proposed for $\mc{N}=1$ gauge theories with fundamental matter the special case $ N_f=1 $ \cite{Bashmakov:2018wts}. Our results allow us to generalize this duality to all values of $N_f$:
\begin{equation}\label{intro_duality}
U(N)_{k+\frac12 (N+N_f),\;k+\frac12 N_f}+N_f\;\Phi \longleftrightarrow SU(k+N_f)_{-N-\frac12 k}+N_f\;\widetilde{\Phi}
\end{equation}
The proposed range of this duality is $k>-\min (N_f,N)$.
The universality discussed above allows us to propose similar $\mc{N}=1$ dualities for other gauge groups. These dualities are:
\begin{align}
SO(N)_{k+\frac12(N-2+N_f)}+N_f\;\Phi&\longleftrightarrow SO(k+N_f)_{-\frac12(k-2)-N}+N_f\;\widetilde{\Phi}\label{intro_SO_duality}\\
Sp(N)_{k+\frac12(N+1+N_f)}+N_f\;\grF&\longleftrightarrow Sp(k+N_f)_{-\frac12(k+1)-N}+N_f\;\widetilde{\grF}\label{intro_Sp_duality}\\
U(N)_{k+\frac12(N+N_f),~k+N+\frac12 N_f}+N_f\;\Phi &\longleftrightarrow U(k+N_f)_{-N-\frac12 k,~-N-k-\frac12 N_f}+N_f\;\widetilde{\Phi}\label{intro_U_duality_1}\\
U(N)_{k+\frac12(N+N_f),~k-N+\frac12 N_f}+N_f\;\Phi &\longleftrightarrow U(k+N_f)_{-N-\frac12 k,~k-N+\frac32 N_f}+N_f\;\widetilde{\Phi}
\label{intro_U_duality_2}
\end{align} 
Note that some of these dualities have already been studied at large $N$ for the case $N_f=1$ \cite{Inbasekar:2015tsa,Jain:2013gza}.

As we shall see, it is of vital importance that to all orders in perturbation theory, there is a single phase transition point $m=m_*$. This fact allows us to match the various phases of these theories exactly across the phase transition point. We emphasize that these dualities are inherently $\mc{N}=1$ supersymmetric, as we do not expect the theories above to have emergent $\mc{N}=2$ SUSY (apart from degenerate cases that we discuss in the text).

The dualities \eqref{intro_SO_duality}-\eqref{intro_U_duality_2} lead to an interesting result for the special case $k=N-N_f$. For example, duality \eqref{intro_U_duality_1} reduces to: 
\beq
U(N)_{\frac32N-\frac12N_f,~2N-\frac12N_f}+N_f\;\grF \longleftrightarrow U(N)_{-\frac32N+\frac12N_f,~-2N+\frac12N_f}+N_f\;\widetilde{\grF}
\eeq{eq:time_reversal_duality}
which means that the theory has emergent time-reversal symmetry in the IR. We obtain a similar result for the rest of the dualities \eqref{intro_SO_duality}-\eqref{intro_U_duality_2}. As discussed in a recent paper, the superpotential in $\mc{N}=1$ time-reversal invariant theories cannot be perturbatively renormalized. Thus, the appearance of a classical moduli space may lead to an exact moduli space in the full quantum theory \cite{Gaiotto:2018yjh}. We thus might expect the theory in \eqref{eq:time_reversal_duality} (and the corresponding theories from the rest of the dualities) to have an exact quantum moduli space. This was shown explicitly for the specific case $N=N_f=1$ of \eqref{eq:time_reversal_duality} by the authors of \cite{Gaiotto:2018yjh} (the emergent time-reversal invariance in this case was also studied in \cite{Benini:2017aed}).

The dualities \eqref{intro_duality}-\eqref{intro_U_duality_2} look very similar to some well-known 2+1$d$ $\mc{N}=2$ dualities \cite{Giveon:2008zn,Aharony:1997bx,Benini:2011mf,Aharony:1997gp}. In fact, we show that most of these dualities can be guessed very naively from known $\mc{N}=2$ dualities, and a discussion of the RG flows of these theories at large $k$ indicate that we can flow from the $\mc{N}=2$ dualities to our $\mc{N}=1$ dualities (some of the dualities above were related more carefully to their $\mc{N}=2$ versions at large $N$ and for $N_f=1$ in \cite{Jain:2013gza}). The only exception is \eqref{intro_U_duality_2}, whose $\mc{N}=2$ version has not yet appeared in the literature. This allows us to conjecture a new $\mc{N}=2$ duality:
\beq 
U(N)_{k+N,~k-N}+N_f\;\Phi \longleftrightarrow U(k+N_f/2)_{-N-k,~k-N+N_f}+N_f \;\widetilde{\Phi}
\eeq{eq:new_N2_duality}
on which we have performed some nontrivial checks. This connection to theories with $\mc{N}=2$ SUSY (in which the superpotential is holomorphic, as opposed to theories with $\mc{N}=1$ SUSY) allows us to better understand the phase diagrams we found. In particular, we discuss how this connection might help explain the unnatural super-critical fixed point at $m=m_*$.

These dualities are also related to the recently proposed non-SUSY bosonization dualities \cite{Aharony:2015mjs,Seiberg:2016gmd,Karch:2016sxi,Benini:2017dus,PhysRevD.94.085009,PhysRevLett.118.011602,Kapustin:2011vz,Armoni:2017jkl,Jensen:2017xbs,Chen:2017lkr,Wang:2017txt,Komargodski:2017keh,Aharony:2016jvv,PhysRevB.95.205137,Aharony:2012nh,Aharony:2011jz,Giombi:2011kc,Cordova:2017vab,PhysRevLett.47.1556,PESKIN1978122,2012arXiv1201.4393B,Son:2015xqa,2015PhRvXo5d1031W,2016PhRvXv6c1026P,2016PhRvBa94x5107W,Murugan:2016zal}. One way in which this can be seen is by naively integrating out the gaugino in the dualities \eqref{intro_duality}-\eqref{intro_U_duality_2} and comparing to the results of \cite{Benini:2017aed,Jensen:2017bjo}. Further evidence for this is that the non-SUSY dualities have been shown to be related to the $\mc{N}=2$ SUSY dualities mentioned above \cite{Jain:2013gza,Gur-Ari:2015pca}. We will not be discussing the relation to the non-SUSY dualities further in this work.

This paper is outlined as follows. We begin by reviewing the relevant background material in Section \ref{background}. In Section \ref{1loop_superpotential_calculation}, we calculate the effective superpotential for an $\mc{N}=1$ gauge theory with matter in an arbitrary representation. Readers interested in the results for the phase diagrams and dualities (and not in the derivation of the effective superpotential) can safely skip most of Section \ref{1loop_superpotential_calculation}, and start at Section \ref{superpotential_summary}. In Section \ref{phase_diagrams} we use this effective superpotential to find the phase diagrams of $U(N)$ and $SU(N)$ gauge theories, and then use symmetry arguments to prove that these phase diagrams are universal. In Section \ref{sec:RG_flows} we discuss the RG flow diagrams of our $\mc{N}=1$ theories when deforming by the available classically marginal $\mc{N}=1$ operators. This allows us to study the emergent symmetries of the CFT at the super-critical point $m=m_*$, and to prove that in generic non-degenerate cases the dualities discussed above are genuine $\mc{N}=1$ dualities. In Section \ref{duality_for_general_Nf} we generalize the $N_f=1$ duality of \cite{Bashmakov:2018wts} to any number of fundamental fields, and conjecture many similar dualities due to the universality of the phase diagram. We also relate most of these dualities to existing $\mc{N}=2$ dualities, and discuss the one exception \eqref{eq:new_N2_duality}.

During the initial stages of this work, some of the main results were also noticed independently by \cite{GomisUnpublished}.

\section{Background}\label{background}
\setcounter{equation}{0}
\subsection{Gauge Theories in \texorpdfstring{$\mc{N}=1$ Superspace}{n1}} \label{gauge_theories_in_superspace}

Our superspace conventions are summarized in Appendix \ref{app_superspace_conventions}. 
The gauge multiplet is described by covariantizing the spinor and vector derivatives:
\beq
\{\na_\al,\na_\be\}=2i\na_\ab~~,~~ \na_\al\equiv D_\al -i \G_\al~~,
\eeq{2.YM}
where the generators are hermitian (which is why there is an $i$ in the definition of the gauge covariant derivative $\na_\al$).
The Bianchi identities imply:
\beq
[\na_\ga,\na_\ab ] = C_{\ga\al}W_\be + C_{\ga\be}W_\al ~~,~~~ \na^\al W_\al  = 0~~.
\eeq{2.Bianchi}
The explicit form of $W_\al$ is
\beq
W_\al = \frac{i}6[\na^\be,\{\na_\be,\na_\al\}]=\frac12 \left( D^\be D_\al \G_\be -i[\G^\be,D_\be\G_\al]-\frac13[\G^\be,\{\G_\be,\G_\al\}] \right)~~.
\eeq{2.Curv}

The Lagrangian for a gauge theory in 2+1$d$ consists of three parts: a Chern-Simons term, a Yang-Mills term, and a matter coupling. 
The CS action is:
\beq
S_{CS}=\frac{k}{2\pi}\int d^2\ta~ \hbox{Tr}\left(\G^\al W_\al+\frac{i}6 \{\G^\al,\G^\be\}D_\al\G_\be+\frac1{12}\{\G^\al,\G^\be\}\{\G_\al,\G_\be\}\right)~~.
\eeq{CS}
Using cyclicity of the trace, this can be written out as:
\beq
S_{CS}=\frac{k}{4\pi}\int d^2\ta~ \hbox{Tr}\left(\G^\al D^\be D_\al \G_\be-\frac{2i}3 \{\G^\al,\G^\be\}D_\al\G_\be-\frac16\{\G^\al,\G^\be\}\{\G_\al,\G_\be\}  \right)~~.
\eeq{CSs}
The YM term is 
\ber
S_{WW}&=&\frac2{g^2}\int d^2\ta ~  \hbox{Tr}~W^2\nonumber\\
&=&\frac1{4g^2}\int d^2\ta~  \hbox{Tr}~( \G^\al D^\be D_\al D^\ga D_\be \G_\ga -2i D^\be D^\al\G_\be[\G^\ga,D_\ga \G_\al]\nonumber\\
&&\qquad\qquad~~- [\G^\be,D_\be \G^\al][\G^\ga,D_\ga \G_\al]-\frac23D^\be D^\al\G_\be[\G^\ga,\{\G_\ga,\G_\al\}]+...)
\eer{WW}
where the higher order terms are not needed for a two-loop calculation. We can rewrite the kinetic terms as:
\beq
L_{kin}=-\frac{k}{4\pi} \, \hbox{Tr}\,(2i\G^\al\pa_\ab\G^\be+\G^\al D_\al D^\be \G_\be ) +\frac1{2g^2} \, \hbox{Tr}\,(\G^\al\square\G_\al -i\G^\al \pa_\ab D^2 \G^\be)
\eeq{Lkin}
When we gauge fix, we also use the identity:
\beq
\hbox{Tr}\,(\G^\al D_\al D^2 D_\be \G^\be)=-\,\hbox{Tr}\,(\G^\al\square\G_\al +i\G^\al \pa_\ab D^2 \G^\be)
\eeq{gfid}
Finally, the gauge invariant matter action is:
\beq
S_{\grF}=\int d^2\ta~ (\frac12\bar \Phi \na^\al \na_\al \Phi +m\bar \Phi \Phi)
\eeq{sxg}

Gauge fixing typically involves nontrivial weighting functions and hence propagating Nielsen-Kallosh ghosts \cite{Nielsen:1978mp,Kallosh:1978de}. Let us consider the naive gauge fixing term (after integrating by parts)
\beq
S_{fix}^0= -\int d^2\ta~\hbox{Tr}\left(\G^\al D_\al
\left(\frac1{2\al g^2}D^2 +\frac{k}{4\pi\be}   \right) D_\be \G^\be\right)
\eeq{fix0}
This gives rise to a (decoupled) free massive Nielsen-Kallosh ghost with a kinetic operator $\left(\frac1{2\al g^2}D^2 +\frac{k}{4\pi\be}   \right) $, 
and a usual Faddeev-Popov ghost action:
\beq
S_{FP}=\frac1{g^2} \int d^2\ta~ \hbox{Tr}\, (BD^\al \na_\al C)
\eeq{sfp}
The gauge-fixed kinetic term is (using \eqref{gfid}):
\ber
L_{kin}&=&-\frac{k}{4\pi} \, \hbox{Tr}\left(2i\G^\al\pa_\ab\G^\be+\left(1-\frac1\be\right)\G^\al D_\al D^\be \G_\be \right) \nonumber\\
&&+\frac1{2g^2} \, \hbox{Tr}\left(\left(1+\frac1\al\right)\G^\al\square\G_\al -i\!\left(1-\frac1\al\right)\G^\al \pa_\ab D^2 \G^\be\right)
\eer{Lkingf}
We use Landau gauge in the following, which is obtained by taking\footnote{It is actually enough to take either $\gra\rightarrow 0$ or $\grb \rightarrow 0$.} $\alpha,\beta\rightarrow 0$. Using \eqref{Lkingf} one can then find the gauge field propagator. Choosing Landau gauge, we find for the gauge field propagator:
\beq
\Delta_\al^{\;\;\be}=-g^2\frac{\delta_\al^{\;\;\be}\left(\kappa D^2+p^2\right)+\left(\kappa-D^2\right)p_\al^{\;\;\be}}{p^2\left(\kappa^2+p^2\right)}
\eeq{gauge_propagator}
where we defined $\kappa=\frac{kg^2}{2\pi}$.

\subsection{Free \texorpdfstring{$\mc{N}=1$}{N=1} Vector Multiplets}
\label{free_N1_vector_multiplets}

We discuss the physics of a free $\mc{N}=1$ vector multiplet. For our purposes, it suffices to consider $\mc{N}=1$ $SU(N)_k$ gauge theories\footnote{The behaviour of $\mc{N}=1$ $U(N)_{k,k'}$ gauge theories can be deduced immediately by noticing that the fields of the vector multiplet are in the adjoint representation, and so their behaviour is affected only by the $SU(N)$ part of $U(N)_{k,k'}$.}. For simplicity, we denote an $\mc{N}=1$ $SU(N)_k$ vector multiplet by $SU(N)_k^{\mc{N}=1}$, and the usual non-SUSY theory by just $SU(N)_k$.

Witten \cite{Witten:1999ds} found that the behaviour of $SU(N)_k^{\mc{N}=1}$ can be split into two regimes. In the small coupling regime $k\geq\frac{N}{2}$, the vacuum is just an $SU(N)_{k-\frac{N}{2}}$ TQFT (this can be seen naively by integrating out the adjoint gaugino, which has a mass proportional to $k$ and with opposite sign). On the other hand, in the strong coupling regime $k<\frac{N}{2}$, the Witten index of these theories vanishes. In this case the vacuum consists of a $U(k+\frac{N}{2})_{-\frac{N}{2}+k,-N}$ TQFT with a Majorana goldstino \cite{Gomis:2017ixy}.
Due to this behaviour, one can write down an $\mc{N}=1$ version of level-rank duality for $SU(N)$ and $U(N)$ vector multiplets:
\begin{equation}\label{eq:N1_level_rank}
U(N)_{k+N/2,k}^{\mc{N}=1}\longleftrightarrow SU(k)_{-N-k/2}^{\mc{N}=1}
\end{equation}
which is valid for $k\geq 0$. Indeed, after integrating out the gaugino we find that this reduces to the usual level-rank duality \cite{naculich1990group, mlawer1991group, Hsin:2016blu}.

The Witten indices of some $\mc{N}=1$ CS theories are collected in Appendix \ref{witten index generalities}.

\subsection{Gauge Theories with a Single Fundamental Matter Multiplet}\label{review_of_adjoint_paper}

The authors of \cite{Bashmakov:2018wts} studied $\mc{N}=1$ $U(N)$ and $SU(N)$ gauge theories with CS terms and with a single fundamental matter multiplet (i.e. $N_f=1$). Due to the structure of the phase diagram, they proposed the following duality:
\begin{equation}\label{eq:Nf1_duality}
U(N)_{k+N/2+1/2,k+1/2}+\Phi \longleftrightarrow SU(k+1)_{-N-k/2}+\widetilde{\Phi}
\end{equation}
where $\grF,\widetilde{\grF}$ are fundamental matter multiplets. In this duality the mass term $\grF^\dagger\grF$ maps to $-\widetilde{\Phi}^\dagger \widetilde{\Phi}$.

This duality was motivated by the study of $\mc{N}=1$ $SU(N)_k$ gauge theories with a massive {\em adjoint} matter multiplet \cite{Bashmakov:2018wts}. For $k\geq 0$, they found that the phase diagrams have three regimes, separated by the values $m=0$ and $m=m_*$ for some $m_*$. For $m<0$ and for $m>m_*$, they found semiclassical vacua which are obtained by integrating out the matter field, leading to $\mc{N}=1$ $SU(N)_{k\pm N/2}$ vector multiplets (the vacuum can then be found using the results of Section \ref{free_N1_vector_multiplets}). Additionally, a new intermediate regime appears in the range $0<m<m_*$. Using the effective 1-loop superpotential, they showed that in this range there are $2^{N-1}$ vacua (in particular, new vacua have appeared from infinity at $m=0$, causing the Witten index to jump). In most of these vacua, the scalar component of $\grF$ condenses, leading to various symmetry breaking patterns for the gauge symmetry. The vacua are thus described by various CS TQFTs. These vacua participate in a series of phase transitions for increasing $m$ up to $m=m_*$,  where the final phase transition occurs, after which we end up with the single vacuum we expected for large positive $m$. (In other words, the phase diagram is similar to the one in Figure \ref{fig:general_phase_diagram}, but now there are many additional phase transitions in the range $0<m<m_*$).

In accordance with this picture, the following phase structure was conjectured for the theories in \eqref{eq:Nf1_duality}, which have a single fundamental matter multiplet instead of an adjoint matter multiplet. The semiclassical phases for large $|m|$ are obtained by integrating out the matter fields, leading to a TQFT in the IR. Again, a new intermediate phase appears for $0<m<m_*$ for some $m_*>0$, which has two vacua. One of these vacua is the same as the vacuum for large negative $m$, while in the other the scalar component of $\grF$ condenses.

We review the behavior around the phase transition point $m_*$ in more detail. On the $U(N)$ side of \eqref{eq:Nf1_duality}, the phases are:
\begin{itemize}
	\item for $m>m_*$ we have one vacuum with a  $U(N)_{k+1}$ TQFT.
	\item for $m<m_*$ we have two vacua, one with a $U(N)_{k}$ TQFT and one with a $U(N-1)_{k+1}$ TQFT.
\end{itemize}
While on the $SU(N)$ side, the phases are:
\begin{itemize}
	\item for $m>m_*$ we have one vacuum with an $SU(k+1)_{-N+1}$ TQFT and one with an $SU(k)_{-N}$ TQFT.
	\item for $m<m_*$ we have one vacuum with an $SU(k+1)_{-N}$ TQFT.
\end{itemize}
Using level-rank duality we find that the phases match around the transition point (after taking $m\rightarrow-m$), and so the duality is consistent.

Note that the duality matches a Higgsed vacuum on one side (e.g. $U(N-1)_{k+1}$) to a vacuum without Higgsing on the other side ($SU(k+1)_{-N+1}$). In addition, note that far away from the transition point (i.e. at $m\ll -g^2$), the phases of the two theories no longer match; on the $U(N)$ side we find a $U(N)_{k}$ TQFT, while on the $SU(N)$ side we find an $SU(k+1)_{-N+1}$ TQFT. This does not affect the duality, since the phases only need to match in the vicinity of the transition point. 
Both of these properties appear in the generalization of the duality to higher $N_f$, which is the focus of this paper.

\begin{figure}[ht]
\centering
\includegraphics[width=0.7\linewidth]{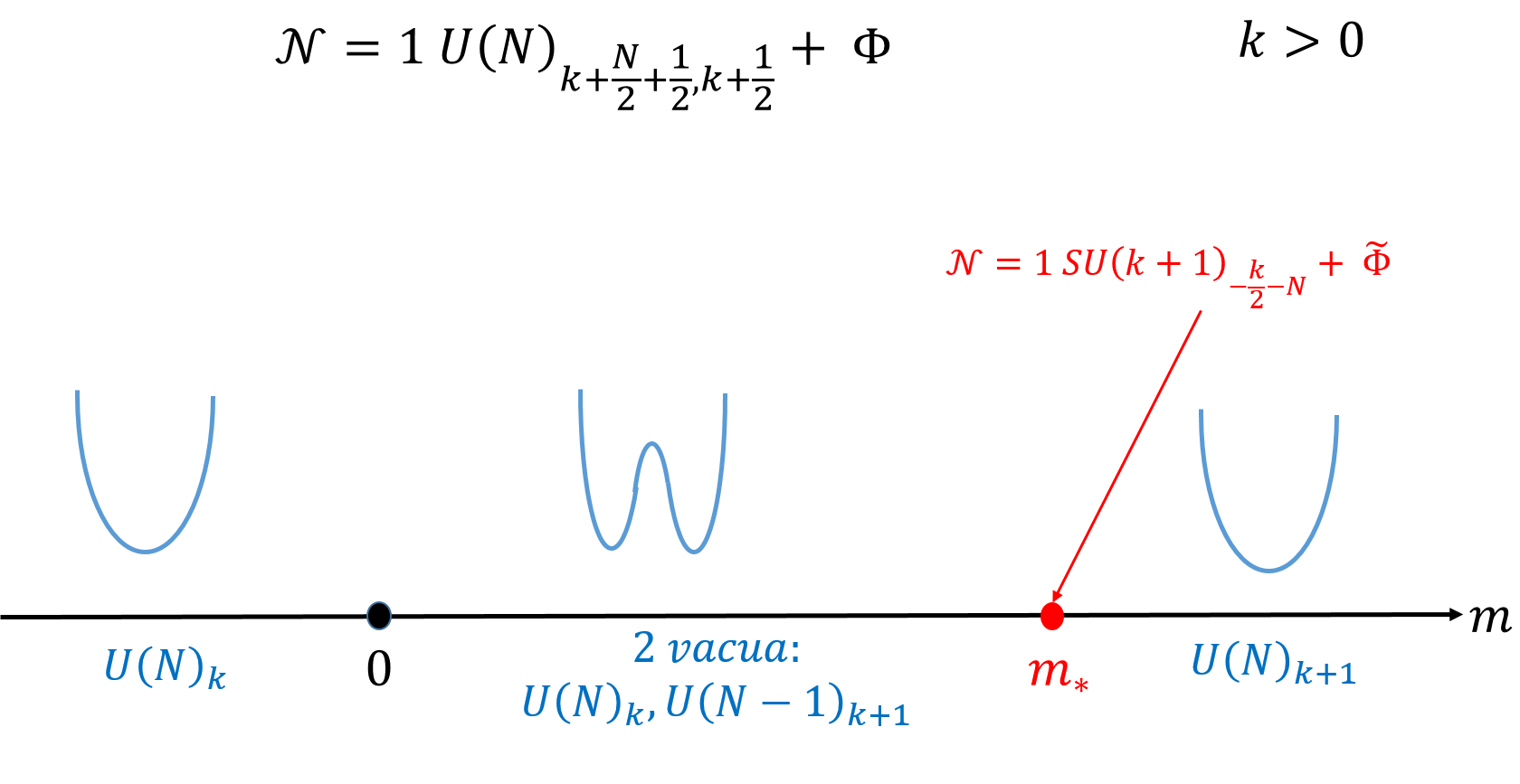}
\caption{\small Phase diagram for an $\mc{N}=1$ $U(N)_{k+\frac{N+1}{2},{k+\frac12}}$ gauge theory coupled to a single fundamental matter multiplet.}
\label{fig:UN_Nf1_diagram}
\end{figure}

A complete picture of the conjectured phases of the $U(N)$ theory with $N_f=1$ appears in Figure \ref{fig:UN_Nf1_diagram}.
In the present paper, we find the 1-loop superpotential for the theories in \eqref{eq:Nf1_duality}, and we prove this phase diagram. We also are able to generalize the phase diagram and the duality to theories with any number $N_f$ of fundamental matter multiplets.

\section{1-Loop Superpotential for Gauge Theories with Arbitrary Matter Representations}\label{1loop_superpotential_calculation}
\setcounter{equation}{0}
\subsection{Comments on the Superpotential and the Potential}\label{comments_on_superpotential}
As usual, the component potential is found by eliminating the auxiliary fields of the scalar multiplets (in three dimensions, the vector multiplet has no auxiliary field).
This leads to an interesting effect; suppose that we expand the superpotential in loops
\beq
W(X,\bar X)=\sum_{\ell=0} W_\ell~.
\eeq{Wloop}
The auxiliary field $F= D^2 X$ enters the bosonic part of the component Lagrangian as:
\beq
L_{aux}=\bar F F+F\frac{\pa W}{\pa X}+\bar F\frac{\pa W}{\pa \bar X}~.
\eeq{LF}
Eliminating $F$ leads to a potential\footnote{This isn't actually completely correct--see \eqref{UZ}.}
\ber
U(X,\bar X)&\!=\!&\frac{\pa W}{\pa X}\frac{\pa W}{\pa \bar X}\equiv U_0+U_1+U_2+...\nonumber\\[2mm]
&=&\frac{\pa W_0}{\pa X}\frac{\pa W_0}{\pa \bar X}+\frac{\pa W_0}{\pa X}\frac{\pa W_1}{\pa \bar X}+\frac{\pa W_1}{\pa X}\frac{\pa W_0}{\pa \bar X}
\nonumber\\[2mm]
&&+\;\frac{\pa W_1}{\pa X}\frac{\pa W_1}{\pa \bar X}+\frac{\pa W_0}{\pa X}\frac{\pa W_2}{\pa \bar X}+\frac{\pa W_2}{\pa X}\frac{\pa W_0}{\pa \bar X}+...
\eer{U}
where the ellipsis indicates contributions beyond two loops. Notice that if the tree-level superpotential vanishes ($W_0=0$), then the {\em two}-loop potential $U_2$ comes 
entirely from the {\em one}-loop superpotential $W_1$, which is a very nice simplification.

Actually, there is another contribution to the potential: in general, the kinetic part of the super-Lagrangian 
$\na^\al \bar X\na_\al X$
gets finite corrections by a multiplicative factor $Z=1+\sum_{\ell=1}Z_\ell$, which leads to a factor in front of the component action
$Z\bar F F$, and hence lead to a component potential
\beq
U(X,\bar X)=\left(Z^{-1}\right)\frac{\pa W}{\pa X}\frac{\pa W}{\pa \bar X}
\eeq{UZ}
Clearly, this does {\em not} modify supersymmetric vacua to any loop level, as they are characterized by $\frac{\pa W}{\pa X}=0$, but could lead to some effects for 
non-supersymmetric vacua. We thus find that in the absence of a tree-level superpotential $W_0$, finding the SUSY vacua of the effective potential $U$ to 2-loop order requires computing $W$ only to 1-loop order.

\subsection{Massless 1-Loop Superpotential}\label{massless_1loop_superpotential}

We can now calculate the 1-loop effective superpotential for $\grF_i$ using the background field method (where $i=1,...,N_f$ is the flavor index). As explained in the introduction, we assume here that all of the "quantum" CS levels are equal. The resulting superpotential is thus applicable to $\mc{N}=1$ gauge theories such as $SU(N)_k$ and $U(N)_{k+\frac12 N,k}$, but for general $U(N)_{k,k'}$ gauge theories an additional calculation is required. We comment on such gauge theories at the end of this section and in Appendix \ref{app_superpotential_UN}.

\begin{figure}[ht]
\centering
\includegraphics[width=0.6\linewidth]{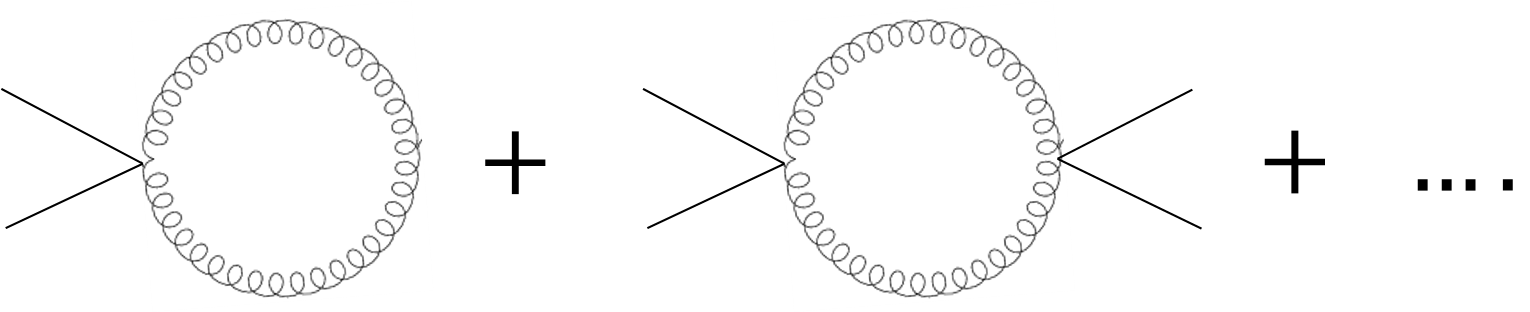}
\caption{\small The 1-loop diagrams we sum over for the 1-loop superpotential}
\label{fig:loops}
\end{figure}

Assume that the matter field $\grF_i$ has a vacuum expectation value (vev) $\grf_i$. Using \eqref{gauge_propagator} we find that in Landau gauge the propagator is transverse:
\beq
\Delta_\al^{\;\;\be} D_\be=0
\eeq{}
This allows us to ignore the vertex
$iD^\gra\Gamma_\gra(\bar\Phi_i\phi_i-\bar\phi_i\Phi_i)$ at one loop in the effective superpotential.\footnote{The vertex actually arises as $-i\Gamma^\gra( D_\gra\bar\Phi_i\phi_i-\bar\phi_i D_\gra\Phi_i)$ because we are only interested in contributions to the effective superpotential and therefore can drop any terms with $ D_\gra \phi$  or $D_\gra\bar\phi$.}
The calculation thus includes only a sum over diagrams with insertions of the vertex $\frac{1}{2}\Gamma^\gra\Gamma_\gra\bar{\phi}_i\phi_i$, depicted in Figure \ref{fig:loops}.

Let us define the superpotential to be\footnote{The action is thus obtained by the usual expression $S=i\int d^3x d^2\theta W$. We note that the 1-loop effective superpotential can be obtained as a direct supersymmetric extension
of the 1-loop Coleman-Weinberg potential:
$$
W=\int \frac{d^3p}{(2\pi)^3}d\theta'
\grd(\theta-\theta')\Tr\log \left(\frac{\grd^2\mc{L}}{\grd{\grF}^2}\right)\grd(\theta'-\theta)
$$} 
$W=\int\frac{d^3p}{(2\pi)^3}d^2\theta'\grd(\theta-\theta')\grS\grd(\theta'-\theta)$. We can thus find $\grS$ by summing over the diagrams:
\begin{align}
\Sigma= & -\frac{1}{2}\mbox{Tr}\frac{\delta_\al^\be(\kappa D^2+p^2)+p_\al^\be(\kappa-D^2)}{p^2\left(\kappa^{2}+p^2\right)}g^2\phi^2\nonumber\\
&+\frac{1}{4}\mbox{Tr}\left(\frac{\delta_\al^\be(\kappa D^2+p^2)+p_\al^\be(\kappa-D^2)}{p^2\left(\kappa^2+p^2\right)}g^2\phi^2\right)^2\nonumber\\
&-\frac{1}{6}\mbox{Tr}\left(\frac{\delta_\al^\be(\kappa D^2+p^2)+p_\al^\be(\kappa-D^2)}{p^2\left(\kappa^2+p^2\right)}g^2\phi^2\right)^3+...
\end{align}
where $\grf^2$ is the $\dim G\times\dim G$ matrix defined by $\grf^2=\bar\grf_i T^{(a} T^{b)}\grf_i$ (with $i=1,..,N_f$ and $a,b=1,...,\dim G$). To simplify this expression we can plug in the following identity:
\beq
\Delta_\al^\ga\Delta_\ga^\be =\frac{2(p^2+\kappa D^2)}{p^2(\kappa^2+p^2)}\Delta_\al^\be
\eeq{}
This allows us to formally sum this series to find
\beq
W=-\int\frac{d^3p}{(2\pi)^3} d^2\theta'\grd(\theta-\theta')\frac{\grd_\gra^\gra}{2}\Tr\log\left(1+\frac{g^2\phi^2(\kappa D^2+p^2)}{p^2(\kappa^2+p^2)}\right)\grd(\theta'-\theta)
\eeq{}
We must take care of the remaining $\theta$ integral. We use the following identities:
\begin{gather}
\grd^2(\theta-\theta')\grd^2(\theta'-\theta)=0\nonumber\\
\grd^2(\theta-\theta')D^\gra\grd^2(\theta'-\theta)=0\nonumber\\
\grd^2(\theta-\theta')D^2\grd^2(\theta'-\theta)=\grd^2(\theta-\theta')
\end{gather}
from which we conclude that $W$ can be rewritten as
\beq
W=-\int\frac{d^3p}{(2\pi)^3}\Tr\log\left(1+\frac{g^2\phi^2(\kappa D^2+p^2)}{p^2(\kappa^2+p^2)}\right)|_{D^2}
\eeq{}
Where $\grS|_{D^2}$ means we are keeping only the terms proportional to $D^2$ after reducing polynomials in $D$ by using the identities from Appendix \ref{app_superspace_conventions}. To calculate this integral, we use the following identity:
\begin{equation}\label{eq:imaginary_trick}
\left(\kappa D^2+p^2\right)^n|_{D^2}=\frac{1}{|p|}\mbox{Im}\left(\left(i\kappa |p|+p^2\right)^n\right)
\end{equation}
We can thus rewrite $W$ as
\begin{equation}\label{eq:log_form}
W=-\int\frac{d^3p}{(2\pi)^3}\frac{1}{|p|}\mbox{Im}\Tr\log\left(1+\frac{g^2\phi^2(i\kappa |p|+p^2)}{p^2(\kappa^2+p^2)}\right)
\end{equation}

We must now evaluate the momentum integral. It is much simpler to calculate $W'$ instead. We find:
\beq
\partial_{g^2}W=
-\mbox{Tr}\int\frac{4\pi p^2dp}{\left(2\pi\right)^3}\frac{\kappa \phi^2}{\left(p^2+g^2\phi^2\right)^2+\kappa^2p^2}=
-\mbox{Tr}\frac{\kappa\phi^2}{2\pi^2}\int dp\frac{p^2}{\left(p^2+g^2\phi^2\right)^2+\kappa^2p^2}
\eeq{}
which results in:
\beq
\partial_{g^2}W =-\mbox{Tr}\frac{\kappa}{4\pi}\frac{\phi^2}{\sqrt{\kappa^2+4g^2\phi^2}}
\eeq{}
and finally
\beq
W =-\frac{\kappa}{8\pi}\mbox{Tr}\sqrt{\kappa^2+4g^2\phi^2}
\eeq{}
reintroducing the color and flavor indices, this can be written as\footnote{Note that as required, the superpotential is time-reversal odd, since in our conventions $k\rightarrow -k$ under time reversal.}
\beq
W =-\frac{\kappa}{8\pi}\mbox{Tr}\sqrt{\kappa^2\grd^{ab}+4g^2\bar\phi_i T^{(a} T^{b)} \grf_i}
\eeq{}
We can now use this result to calculate the 2-loop effective potential for $\grf$ (as explained in Section \ref{comments_on_superpotential}). We note that as explained above, since the tree-level superpotential vanishes, the solutions of the F-term equations due to this superpotential are actually the 2-loop solutions when written in components.

\subsection{Adding a Mass Term}

We now add a mass to the matter field $\Phi$, which amounts to adding a tree-level superpotential of the form  $ W_0=m\bar{\Phi}\Phi $. The effective superpotential now takes the form
\beq
W= m\grf_i^2-\frac{\kappa}{8\pi}\Tr \sqrt{\kappa^2\delta^{ab}+4g^2\bar\phi_i T^{(a} T^{b)} \phi_i} 
\eeq{}
The important point here is that adding a mass term did not change the 1-loop contribution from the previous section, since we do not have any matter fields running in the loops in Figure \ref{fig:loops}. This means that the massive result is just the sum of the tree level and the massless 1-loop result (of course, at higher loop orders this will not work).
Again, one can now calculate the effective potential using the methods described in Section \ref{comments_on_superpotential}.

\subsection{Summary}\label{superpotential_summary}

We have thus found that the 1-loop effective superpotential for massive matter in an arbitrary representation is
\begin{equation}\label{eq:full_superpotential}
W= m|\grF_i|^2-\frac{\kappa}{8\pi}\Tr \sqrt{\kappa^2\delta^{ab}+4g^2\bar\Phi_i T^{(a} T^{b)} \Phi_i}
\end{equation} 
We can now use this to find the SUSY vacua by solving the F-term equations $W'=0$. We emphasize that when $m=0$, the resulting solutions are the correct vacua to 2-loop order in the component fields. To simplify notation in the following, we redefine the fields and the couplings so that the effective superpotential $W$ is proportional to
\begin{equation}\label{eq:superpotential}
\widetilde{W}= m|\grF_i|^2-\Tr \sqrt{\delta^{ab}+\bar\Phi_i T^{(a} T^{b)} \Phi_i} 
\end{equation}
Furthermore, since we are only interested in the solutions to the F-term equations in the following, we can disregard the proportionality constant and treat $\widetilde{W}$ in \eqref{eq:superpotential} as our superpotential.

We can write down the superpotential \eqref{eq:superpotential} explicitly for specific representations and gauge groups. First, we find the superpotential for a $U(1)_k$ gauge theory with $N_f$ charge 1 fields. Assuming the matter fields have vevs $\grf_i$, the superpotential reduces to the simple form
\begin{equation}\label{eq:superpotential_U1}
\widetilde{W}_{U(1)}= m|\grf_i|^2-\sqrt{1+|\grf_i|^2}
\end{equation}

Next, we can discuss the adjoint representation of $SU(N)$ and $U(N)$. Assume that $\grf$ gets a constant vev. For $\grf$ in the adjoint representation of $U(N)$ or $SU(N)$, we can think of $\grf$ as an $N\times N$ matrix. We can then use gauge transformations to put the vev matrix $\grf$ in the form $\grf=\text{diag}(\grf_1,...,\grf_N)$ (remembering that for $SU(N)$ we have $\sum_i\grf_i=0$). Plugging this into the superpotential \eqref{eq:superpotential} we obtain
\begin{equation}
\widetilde{W}_{\text{adj}}= m\grf_i^2-\sum_{i,j=1}^N\sqrt{1+(\grf_i-\grf_j)^2}
\end{equation} 
The resulting effective potential can be compared to those obtained in \cite{Armoni:2005sp,Armoni:2006ee}.

Next, we discuss $SU(N)$ and $U(N)$ gauge theories with fundamental matter (where as explained above, the two quantum CS levels for $U(N)$ are identical). We can try to repeat the process described above. Assuming we have $N_f$ fields in the fundamental representation, we can think of $\grf$ as an $N\times N_f$ matrix. Using combined gauge and flavor transformations, we can then choose a representative of the vev matrix $\grf$ which is in upper-diagonal form. Explicitly, for $N_f\leq N$ we have 
\beq 
\grf=\begin{pmatrix} \grf_1\\&...\\& & \grf_{N_f}\\
0 & ...& 0\\
\vdots &  & \vdots\\
0  & ...& 0
\end{pmatrix}
\eeq{}
with $\grf_i>0$. We have a similar form for $N_f> N$ (with $N$ terms on the diagonal).
However, this still does not give a simple form for the superpotential for an $SU(N)$ or a 
$U(N)$ gauge theory with fundamental matter (so that we have to restrict ourselves to a numerical study of this superpotential). One case in which the superpotential is simplified is for an $SU(N)_{k+\frac12 N}\times U(1)_k$ gauge theory with fundamental matter. For $N_f> N$ we obtain
\begin{equation}\label{eq:fund_superpotential}
\widetilde{W}_{SU(N)_{k+\frac12 N}
\times U(1)_k}= m\grf_i^2-\sum_{i,j=1}^{N}\sqrt{1+\grf_i^2+\grf_j^2}
\end{equation} 
For $N_f\leq N$, if we take the vector $\phi_i$ of length $N_f$ and pad it with zeros to make it of length $N$, then the result is identical to \eqref{eq:fund_superpotential}\footnote{More generally, for both $N_f\leq N$ and $N_f>N$ we can write a $U(N_f)$ invariant expression for $W$ as well:
$$
W= m\tr Q+2(N-N_f)\Tr\sqrt{1+Q}+(\Tr\otimes\Tr)\sqrt{1+Q\otimes 1+1\otimes Q}
$$ where $Q_\gra^\grb$ is the field bilinear $Q_\gra^\grb=\grf_\gra^*\grf^\grb$.}.

Next we discuss $SO(N)$ gauge theories with fundamental matter. The idea is the same as before. After putting $\grf$ into upper-diagonal form, we find for $N_f>N$
\begin{equation}\label{eq:fund_superpotential_SO}
\widetilde{W}_{SO(N)_k}= m\grf_i^2-\sum_{i>j}^{N}\sqrt{1+\grf_i^2+\grf_j^2}
\end{equation} 
And again, this expression is also true for $N_f\leq N$ if we pad $\grf_i$ with zeros to make it of length $N$.

Finally, we must discuss a general $\mc{N}=1$ $U(N)_{k,k'}$ gauge theory. When the quantum CS levels for the fields are different, the calculation above becomes much more complicated. In this case, we must restrict ourselves to an integral form for the superpotential, similar to the form \eqref{eq:log_form}, which allows us to perform a numerical analysis. We describe the calculation in Appendix \ref{app_superpotential_UN}.

\section{Phase Diagrams}\label{phase_diagrams}
\setcounter{equation}{0}
We can now use the 1-loop superpotential \eqref{eq:superpotential} to find the phase structure of $\mc{N}=1$ gauge theories with fundamental matter. To do this, we give an arbitrary vev to the matter fields $\grf_i$, and find the SUSY vacua by solving the F-term equations $\frac{\partial W}{\partial \phi_i}=0$. We start by performing this analysis for two cases in which the superpotential is particularly simple: a $U(1)_k$ gauge theory and an $SU(N)_{k+\frac12 N}\times U(1)_k$ gauge theory. What we find is that the behavior of the $SU(N)_{k+\frac12 N}\times U(1)_k$ gauge theory is universal, and appears also in the $SU(N)_k$ and $U(N)_{k+\frac{N}{2},k}$ gauge theories (the resulting phase diagrams are of the form discussed in Figure \ref{fig:general_phase_diagram}). Note that when the gauge group is $SU(2)$, the theory requires special treatment. In general, an $SU(N)$ gauge theory with $N_f$ fundamentals has $U(N_f)$ global symmetry. But in an $SU(2)$ gauge theory, this symmetry is classically enhanced to $Sp(N_f)$. The phases of an $SU(2)$ gauge theory are discussed separately in Section \ref{su(2) phase}.

We find that all of these theories have the same general form for their phase diagram. In Section \ref{symmetry_argument} we use a symmetry argument to explain why this is the case. This argument also allows us to prove that there is a single phase transition point to all orders in perturbation theory.

The main results of this section are summarized in Section \ref{sec:phases_summary}. Using the results of this section, we are able to extend the duality proposed in \cite{Bashmakov:2018wts} to larger $N_f$ in the next section.

\subsection{Warmup: Phases of \texorpdfstring{$U(1)_{k+\frac{N_f}2}$}{U(1)k}
with \texorpdfstring{$N_f$}{Nf} Matter Multiplets}

We look for the vacua of an $\mathcal{N}=1$ $U(1)_{k+\frac{N_f}2}$ vector multiplet coupled to $N_f$ charge 1 fields. We give the fields a constant vev, and plug them into the superpotential found in equation \eqref{eq:superpotential_U1}:
\beq
\widetilde{W}_{U(1)}= m|\grf_i|^2-\sqrt{1+|\grf_i|^2}
\eeq{}
We look for the solutions to the F-term equations $W'=0$. We find three different regimes:
\begin{itemize}
\item For $m<0$ we find a single solution, $\grf_i=0$. The vacuum in this case is $U(1)_{k}$.

\item For $0<m<m_*$ (for some $m_*>0$) we find two solutions: one at $\grf_i=0$ (with low-energy theory $U(1)_{k}$) and one at $|\grf_i|^2=\frac{1-4m^2}{4m^2}$. Let us discuss the vacuum at the latter solution. First, we note that the solution runs off to infinity when $m\rightarrow 0$. Second, we find that it breaks the gauge symmetry and the flavor symmetry. The gauge symmetry is broken to nothing, but the breaking of the flavor symmetry leaves us with a NLSM with target space $\mathbb{CP}^{N_f-1}=\frac{U(N_f)}{U(N_f-1)\times U(1)}$ in the IR. So the low-energy theory is a $\mathbb{CP}^{N_f-1}$ NLSM.

\item Finally, for $m>m_*$ we once again find the single solution $\grf_i=0$, this time with low-energy theory $U(1)_{k+N_f}$. 
\end{itemize}

Let us discuss the Witten index of the theory. As we saw, for small positive $m$ a new vacuum appears from infinity. This is due to the changing of the asymptotics of the effective potential (as discussed in \cite{Bashmakov:2018wts}), and can cause the Witten index to jump \cite{Witten:1982df}. Indeed, we find that the phases above have the following Witten indices (see Appendix \ref{app_matching_witten_index}):
\begin{center}
\begin{tabular}{c  c c c }
 & $m<0$ & $0<m<m_*$ & $m>m_*$ \\ 
\hline
Witten Index & $k$ & $k+N_f$ & $k+N_f$   
\end{tabular}
\end{center}
And so the Witten index jumps only at $m=0$, as expected.

Another interesting point is $m=m_*$. There, we find a phase transition where two vacua merge into one. This must be a phase transition of second order or higher, since these are SUSY vacua which have zero energy. We describe the dual theory at this phase transition later on.

\begin{figure}[ht]
\centering
\includegraphics[width=0.7\linewidth]{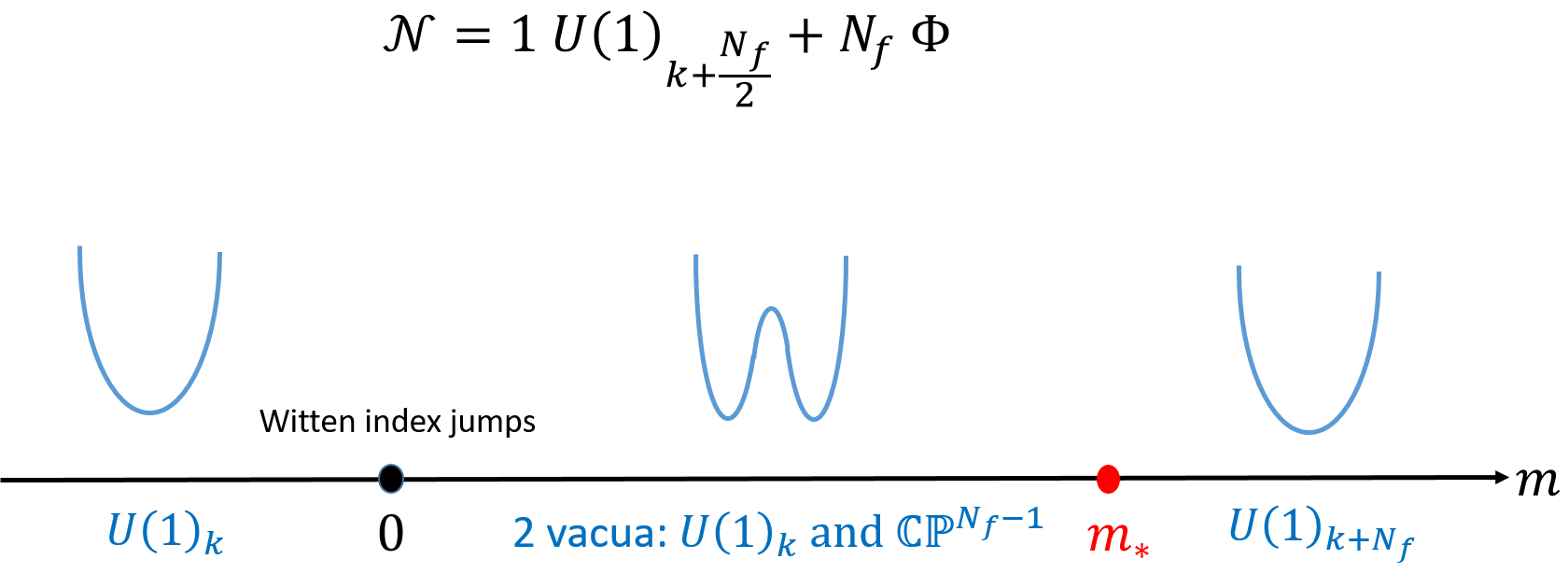}
\caption{\small Phase diagram for an $\mc{N}=1$ $U(1)_{k+\frac{N_f}{2}}$ gauge theory coupled to $N_f$ charge 1 matter multiplets.}
\label{fig:U1_Nf_diagram}
\end{figure}

This discussion is summarized in Figure \ref{fig:U1_Nf_diagram}.

\subsection{More Warmups: Vacua of \texorpdfstring{$SU(N)_{k+\frac12 N}\times U(1)_k$}{SUNk} 
with \texorpdfstring{$N_f$}{Nf2} Fundamentals}
\label{phases_of_SUN_times_U1}

Before attempting to find the phases for $SU(N)$ and $U(N)$ gauge theories, we look for the phases of $SU(N)_{k+\frac12 N}\times U(1)_k$. The reason we do this is that (as we saw in equation \eqref{eq:fund_superpotential}) the superpotential for this case simplifies. Thus, we can study the solutions to the F-term equations analytically for this case. Unfortunately, we could not find such a simple form for $SU(N)$ and $U(N)$ gauge theories, and so we have to resort to numerical calculations to study them in following sections. However, since we show that the phase diagrams of all of these theories are similar, it is worthwhile to exhibit an example where we can find the solutions analytically.

The 1-loop effective superpotential for a $SU(N)_{k+\frac12 N}\times U(1)_k$ theory is:
\beq
\widetilde{W}= m\sum_i \grf_i^2-\sum_{i,j=1}^{N_f}\sqrt{1+\grf_i^2+\grf_j^2} 
\eeq{eq:temp_superpotential_SUNtimesU1}
where $\grf_i$ are the eigenvalues after using gauge and flavor transformations to bring $\grf$ to an upper-diagonal form. 
Let us discuss the solutions to the F-term equations $\frac{\partial}{\partial\phi_i}\widetilde{W}=0$.
We find that the solution $\grf_0=0$ exists for any $m$. Additionally, for a specific range $0<m\leq m_*$, we have $\min(N,N_f)$ additional solutions $\grf_n$ of the form\footnote{we are abusing notation here, since $\grf_n$ is an $N\times N_f$ rectangular matrix. The point is that $\grf_n$ is upper-diagonal with $n$ identical values on the diagonal, with the rest of the values on the diagonal vanishing. This explains why there are $\min(N,N_f)$ solutions of this form. We continue to use this notation throughout this paper.} 
$$ \grf_n=\begin{pmatrix}
v_n(m)\cdot \id_{n\times n} & \\
& 0
\end{pmatrix} $$
where $v_n\in\mathbb{R}$ and  $\id_{n\times n}$ is the $n\times n$ unit matrix.

Explicitly, we find that in this case, $m_*=N$ (in the units used in equation \eqref{eq:temp_superpotential_SUNtimesU1}).
We also find $\lim_{m\rightarrow 0^+}v_n=\infty$, and so we learn that these solutions come in from infinity at $m=0$. This is related to the fact that the asymptotics of the effective potential change at $m=0$, which as we see leads to a jump in the Witten index. Furthermore, we find that all of these solutions coalesce simultaneously at the phase transition point $m=m_*$. In other words, we find that all $v_n$ vanish at $m=m_*$ simultaneously at 1-loop. The phase diagram thus consists of a single point where the Witten index jumps, and a single phase transition point $m_*$. We thus find a phase diagram of the form discussed in Figure \ref{fig:general_phase_diagram}.

We will not study the specific vacua and phases in detail, since we are more interested in the results for $SU(N)$ and $U(N)$ gauge theories. But what we find is that the solutions $\grf_n$ play a pivotal role in the following sections. Their appearance here is not a coincidence, and we show that they are also the form of the solutions for the 1-loop F-term equations for the $U(N)$ and $SU(N)$ gauge theories. In fact, the picture described here is almost identical to the picture in the $U(N)$ and $SU(N)$ case - these vacua appear from infinity at $m=0$, causing a jump in the Witten index, and coalesce at a single phase transition point at some $m=m_*$. Later we explain why this behavior is universal using symmetry arguments. These arguments allow us to show that the $\grf_n$ solve the F-term equations to all orders in perturbation theory, from which we learn that there is a single phase transition point to all orders in perturbation theory.

\subsection{Phases of \texorpdfstring{$SU(N)_{k+\frac12 N}$}{SUN2} with \texorpdfstring{$ N_f $}{Nf3} Fundamentals}
\label{phases_of_SUN}

\begin{figure}[ht]
\centering
\includegraphics[width=0.8\linewidth]{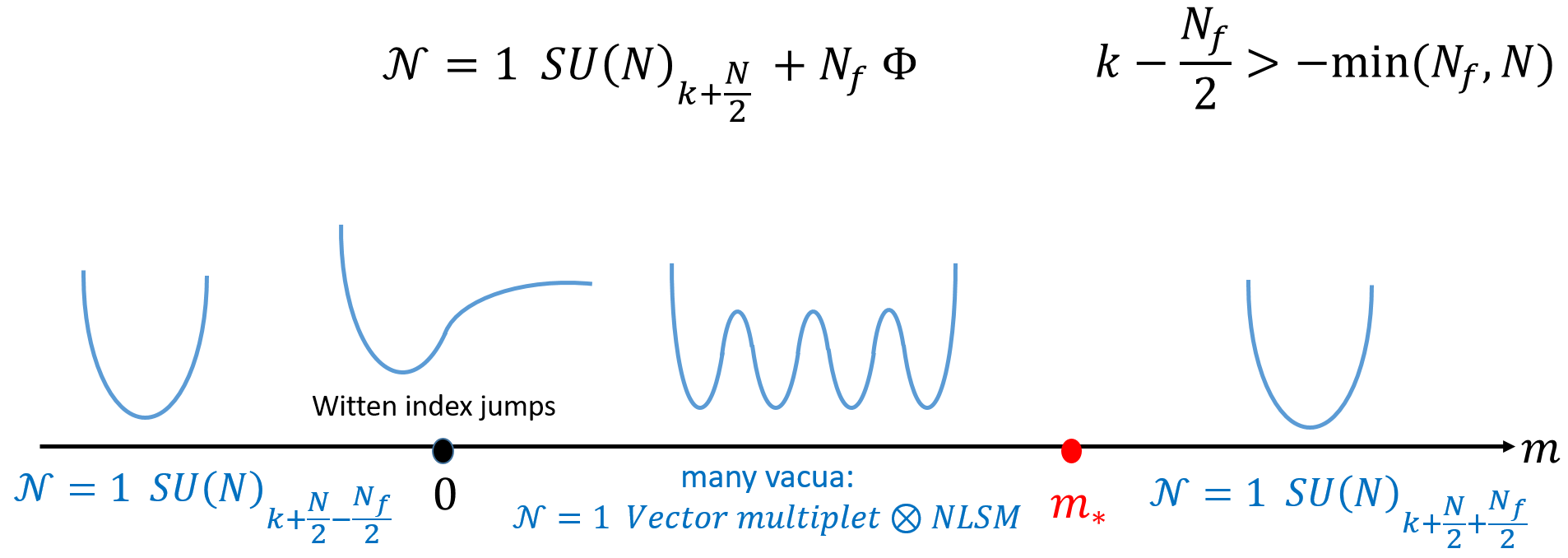}
\caption{\small Phase diagram for an $\mc{N}=1$ $SU(N)_{k+\frac12 N,k}$ gauge theory coupled to a $N_f$ fundamental matter multiplets. The precise form of the low-energy theories in the intermediate phase is discussed in the text.}
\label{fig:SUN_diagram}
\end{figure}

We now study the phase diagrams for an $\mc{N}=1$ $SU(N)_{k+\frac12 N}$ gauge theory with $N_f$ fundamentals\footnote{Due to our conventions of the CS level, it suffices to consider $k+\frac{N}{2}\geq0$.}. 
The result is very similar to the $SU(N)\times U(1)$ case studied in the previous section. In particular, the form of the phase diagram is the same as in Figure \ref{fig:general_phase_diagram}, and most of this section focuses on finding the different vacua in the intermediate phase. A more detailed form of our resulting phase diagram appears in Figure \ref{fig:SUN_diagram}.

Let us find the SUSY vacua of the theory. We start by putting the vev matrix in upper-diagonal form. Plugging this form into the superpotential from Section \ref{superpotential_summary}, we can solve the F-term equations numerically as a function of $m$ (as discussed in Section \ref{superpotential_summary}, it is difficult to find the solutions to the F-term equations for $SU(N)$ analytically, and so we restrict ourselves to numerical results). We find that the solutions are identical to the ones in the $SU(N)\times U(1)$ case discussed in Section \ref{phases_of_SUN_times_U1}. Specifically, we find that there are three distinct regions as a function of $m$. For large $|m|$, the vacuum is just the result of integrating out the massive matter, so that the vacuum is $SU(N)_{k+\frac12 (N\pm N_f)}$. However, in some range $0<m< m_*$, we find $\min(N,N_f)$ new solutions of the form $ \grf_n=\begin{pmatrix}
v_n(m)\cdot \id_{n\times n} & \\
& 0
\end{pmatrix} $. Specifically, we find once again that there is only one phase transition at $m=m_*$ (and a jump in the Witten index at $m=0$).

We can now find the explicit form of the low-energy theories at the solutions $\grf_n$. We study the two cases $N_f<N$ and $N_f\geq N$ independently.

\subsubsection{\texorpdfstring{$N_f < N$}{Nf<}}\label{SUN_Nf_smaller_N}
We find the phases for $N_f\leq N$ with small positive mass. As explained above, we have a SUSY vacuum for each of the solutions $\grf_n$ for $n=0,..,N_f$. Let us consider a specific solution $\grf_n$. In the $\grf_n$ vacuum, both the gauge symmetry and the global symmetry are broken. Higgsing the gauge group leads to an $SU(N-n)$ vector multiplet, while the breaking of the flavor symmetry leads to to a $\frac{SU(N_f)}{S[U(n)\times U(N-n)]}$ NLSM (we denote this NLSM by $\mc{M}_{N_f,n}$). 
The vacua corresponding to the solutions $\grf_n$ are then:    
\begin{alignat}{4}
      \phi_0:\qquad && &SU(N)_{k+\frac12 (N-N_f)}^{\mc{N}=1}\nonumber\\
      \phi_1:\qquad && \mc{M}_{N_f,1}\times & SU(N-1)_{k+\frac12 (N-N_f+1)}^{\mc{N}=1}\nonumber\\
      &&...&\nonumber\\
      \phi_n:\qquad && \mc{M}_{N_f,n}\times & SU(N-n)_{k+\frac12 (N-N_f+n)}^{\mc{N}=1}\nonumber\\
      &&...&\nonumber\\
      \phi_{N_f}:\qquad && &SU(N-N_f)_{k+\frac12 N}^{\mc{N}=1}
\end{alignat}
where we have emphasized that the vacua are $\mc{N}=1$ vector multiplets. It is easy to understand the vacua above; at the solution $\grf_n$, $n$ multiplets have a non-zero vev, while the $N_f-n$ remaining multiplets are massive. We thus break the gauge group to $N-n$ and shift the CS level by $\frac{N_f-n}{2}$. The vacua described above are not the final solutions. To find their precise form, we must take special care of the resulting $\mc{N}=1$ vector multiplets, as described in Section \ref{free_N1_vector_multiplets}. Specifically, for $k\geq \frac{N_f}{2}$ they all result in TQFTs, while for $k<\frac{N_f}{2}$ some solutions break SUSY dynamically.

We can also study the behavior of the Witten index of the theory. In Appendix \ref{app_SUN_N>Nf}, We show that the Witten index jumps only at $m=0$, and does not jump at the phase transition point $m=m_*$ (this is similar to the result in \cite{Bashmakov:2018wts}). The picture is thus the following. For negative mass, there is one vacuum. At small positive mass, $N_f$ new vacua appear from infinity (which lead to the jump in the Witten index). These vacua merge together with the vacuum at the origin at the phase transition point $m=m_*$, resulting in a single vacuum for $m>m_*$.

In Section \ref{symmetry_argument} we prove that this picture persists to all orders in perturbation theory. In particular, the solutions to the F-term equations of the full superpotential also coalesce at a single point, so that there is only one phase transition.

\subsubsection{\texorpdfstring{$N_f\geq N$}{N>}}
The case $N_f\geq N$ has a few additional subtleties that we must take care of. Specifically, in this case there is a vacuum $\grf_N$ in which the gauge group is completely Higgsed, and we find that the solutions "truncate" when we reach this maximally Higgsed solution. Furthermore, in this solution the baryons (defined in the usual way $Q^{i_1...i_{N}}=
\epsilon^{a_1...a_{N}}
\phi_{a_1}^{i_1}...\phi_{a_{N}}^{i_N}$) get a vev. This changes the symmetry breaking pattern, since the baryons are charged under the flavor symmetry, and so we must be careful about the resulting NLSM.

We find that the vacua for small positive mass are:
\begin{alignat}{4}
      \phi_n:\qquad && & \mc{M}_{N_f,n}\times SU(N-n)_{k+\frac12 N-\frac12(N_f-n)}^{\mc{N}=1} \nonumber\\
      \phi_N:\qquad && &\frac{U(N_f)}{SU(N)\times U(N_f-N)} 
\end{alignat}
where $n=0,...,N-1$. Once again, we find that the Witten index jumps only at $m=0$ (and not at the phase transition point). This is proven in Appendix \ref{app_SUN_N<Nf}.

\subsection{Phases of \texorpdfstring{$U(N)_{k+\frac12 N,k}$}{UNk} with \texorpdfstring{$ N_f $}{Nfa} Fundamentals}
\label{phases_of_UN}

\begin{figure}[h]
\centering
\includegraphics[width=0.8\linewidth]{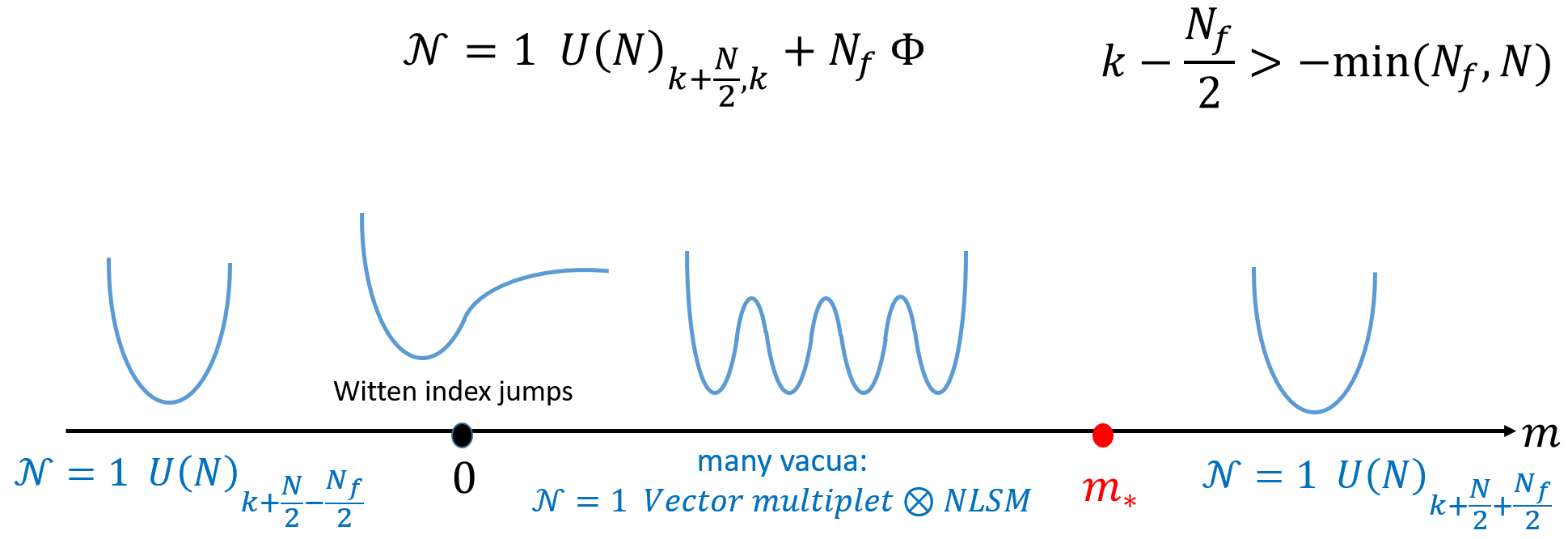}
\caption{\small Phase diagram for an $\mc{N}=1$ $U(N)_{k+\frac21 (N+N_f),k+\frac12N_f}$ gauge theory coupled to a $N_f$ fundamental matter multiplets. The precise form of the vacua in the intermediate phase is discussed in the text.}
\label{fig:UN_diagram}
\end{figure}

We now study the phase structure of $U(N)_{k+\frac12 N,k}$ gauge theories as a function of the mass $m$ of the matter multiplets. Again we give all of the $N_f$ matter multiplets the same mass $m$. One immediately finds that for large $|m|$, the vacuum is just the result of integrating out the massive matter multiplets, so that the large mass phases are $\mc{N}=1$ $U(N)_{k+\frac{N\pm N_f}{2},k\pm \frac{N_f}{2}}$ vector multiplets. However, for $0<m< m_*$, the solutions $\grf_n$ appear once again, and a more careful analysis must be done. We perform this analysis separately for $N_f\leq N$ and $N_f> N$. The resulting phase diagram appears in Figure \ref{fig:UN_diagram}
(note that this phase diagram is still of the general form discussed around Figure \ref{fig:general_phase_diagram}).

\subsubsection{\texorpdfstring{$N_f\leq N$}{Nf>>}}
Let us discuss the vacuum at each $\grf_n$. Once again, denote by $ \mathcal{M}_{N,n} $ a NLSM with target space $ \frac{U(N)}{U(N-n)\times U(n)} $. The vacua of the theory for small positive mass are:
\begin{alignat}{4}
      \phi_n:\qquad && & \mc{M}_{N_f,n}\times U(N-n)_{k+\frac12 N-\frac12(N_f-n),~k-\frac12 N_f+n}^{\mc{N}=1}
\end{alignat}
With $n=0,...,N_f$. As we show in Appendix \ref{app_UN_N>Nf}, the Witten index jumps only at the point $m=0$.

The shift of the $U(1)$ CS level must be explained. In addition to the shift due to integrating out the matter multiplets, there appears to be another shift of the $U(1)$ CS level. This is due to the fact that by definition, a $U(N)_{K,K'}$ CS theory has a Lagrangian of the form
\beq
\frac{K}{4\pi}\Tr (A\wedge d A-\frac{2i}{3}A\wedge A\wedge A) + \frac{K'-K}{4\pi N}\Tr A\wedge\Tr d A
\eeq{}
and so Higgsing the gauge group also shifts the value of the $U(1)$ CS level\footnote{Note that such an additional shift must occur since a $U(N)_{K,K'}$ TQFT is consistent only when $K'-K\equiv 0\mod N$ \cite{Hsin:2016blu}} $K'$.

\subsubsection{\texorpdfstring{$N_f> N$}{Nf>>>}}
Once again, we find a similar result for $N_f>N$ with some minor changes. We again find that the phases truncate when we reach maximal Higgsing at the solution $\grf_N$, so that the phases for small positive mass are:
\begin{alignat}{4}
      \phi_n:\qquad && & \mc{M}_{N_f,n}\times U(N-n)_{k+\frac12 N-\frac12(N_f-n),~k-\frac12 N_f+n}^{\mc{N}=1}
\end{alignat}
with $n=0,...,N$. Once again, to find the exact form of the vacua one must be careful about the value of $k$, and whether or not SUSY is broken. In Appendix \ref{app_UN_N<Nf} we show that once again the index jumps only at $m=0$.

\subsection{\texorpdfstring{$SU(2)$}{su2} Gauge Symmetry and Enhanced 
Global Symmetry}
\label{su(2) phase}

We now discuss the case of an $SU(2)$ gauge theory. This case is special, since the global symmetry is enhanced at the classical level from $U(N_f)$ to $Sp(N_f)$. This leads to a different vacuum structure in the phases discussed above, since the global symmetry breaking pattern is different (and so the NLSM is different). However, we show that the discussions that appeared above (and in particular the matching of the Witten index across the phase transition) applies here as well.
  
We start by explicitly describing the enhancement of the global symmetry in a theory with an $SU(2)$ gauge theory coupled to a boson in the fundamental representation. The generalization to a SUSY theory is immediate.
The main point is that for an $SU(2)$ doublet $\grf$, the Lagrangian
 \begin{align*}
\mathcal{L}=|D_\mu \phi_i|^2 + m |\phi_i|^2,~~~
D_\mu = \partial_\mu -iA_\mu
\end{align*}
can be re-written as a function of 
$
\tilde{\phi}^a\equiv
(
\phi_1^a,...,\phi^a_{N_f},
\epsilon^{ab}\phi_{1b}^*,...,\epsilon^{ab}\phi_{N_fb}^*
)$, where $a,b$ are color indices. The resulting Lagrangian is
\begin{align*}
\mathcal{L}=\frac{1}{2}\Omega_{IK}(D_\mu  \tilde{\phi}_I) \epsilon (D^\mu\tilde{\phi}_K) + \frac{m}{2} \Omega_{IK}\tilde{\phi}_{I}\epsilon\tilde{\phi}_{K},~~~~
\Omega=\begin{pmatrix}
0&\id_{N_f}\\-\id_{N_f}&0
\end{pmatrix}
\end{align*}
here, the color indices are implicitly contracted using $\epsilon$ and $I,K=1,...,2N_f$. One can check that this Lagrangian is invariant under an $Sp(N_f)$ transformation
\begin{equation}
\begin{aligned}
\tilde{\phi}_I^a \longrightarrow S_{IK}\tilde{\phi}_K^a, ~~~~
&S^t\Omega S=\Omega
\end{aligned}
\end{equation}

Let us discuss global symmetry breaking pattern in this model. The moduli space can be parametrized by the gauge invariant building blocks
$$
\mathcal{M}^I_{J}=\tilde{\phi}^{\dagger Ia} \tilde{\phi}_{Ja}
$$ 
Now, for any $N_f$, we can use gauge and global symmetry transformations on $\grf$ to make $\mc{M}^I_J$ vanish apart from one component $\mc{M}^1_{1}=\mc{M}^{N_f+1}_{N_f+1}=v$. If $v=0$ then there is no symmetry breaking, while for any $v\neq 0$ the symmetry breaking pattern results in a NLSM with target space
\begin{align*}
\frac{Sp(N_f)}{Sp(1)\times Sp(N_f-1)}
\end{align*}

We can now find the phase diagram of
an $\mc{N}=1$ $SU(2)_{k+1+\frac12 N_f}$ gauge theory with any number $N_f\geq1$ of fundamentals.
As for the theories above, the phase diagram as a function of $m$ consists of three regions. For $m<0$ we get an $SU(2)_{k+1}$ $\mathcal{N}=1$ vector multiplet. For $0<m<m_*$ we get two vacua, one with an $SU(2)_{k+1}$ vector multiplet and one with an $\frac{Sp(N_f)}{Sp(1)\times Sp(N_f-1)}$ NLSM. Finally, for $m>m_*$ we find an $SU(2)_{k+1+N_f}$ $\mathcal{N}=1$ vector multiplet. Note that the Witten index still matches across the phase transition at $m=m_*$, since the Euler characteristic of the target space is given by $\chi(\frac{Sp(N_f)}{Sp(1)\times Sp(N_f-1)})=N_f$.

\subsection{Universality of Solutions and the Phase Transition}\label{symmetry_argument}

The theories we have discussed exhibit a very interesting pattern. It seems as though the general form of the solutions to the F-term equations at 1-loop does not depend on the the CS level or the gauge group. In other words, the form of the solutions $\grf_n=\begin{pmatrix}
v_n\cdot \id_{n\times n} & \\
& 0
\end{pmatrix}$ described above seems to be universal for gauge theories with fundamental matter (although the specific values of the $v_n$'s are not universal). Additionally, it is interesting that at one loop there is a single phase transition point, even when the number of vacua is large.

We now use a symmetry argument to explain both of these properties. First we show that these solutions are indeed universal in a certain sense, and that they appear to all orders in perturbation theory. This also allows us to show that there is a single phase transition to all orders in perturbation theory. This result is very important for our conjectured dualities, since the fact that there is a single phase transition means that we must match all of the vacua on both sides of the duality. We prove these properties for $U(N)$ and $SU(N)$ gauge theories with fundamental matter, but as we comment below, the proof can easily be applied to more gauge groups with fundamental matter.

We first study the symmetries of the superpotential.
A general solution of the F-term equations $\grf$ can be put in upper diagonal form using combined gauge and flavor rotations. Let us label the resulting values on the diagonal by $v_1,...,v_K$ (where $v_i\in\mathbb{R}$ and $K=\min(N,N_f)$). It is crucially important to find the residual symmetries of the solution $\grf$. First, we can permute the different $v_i$'s, and so we have an $S_K$ symmetry. Furthermore, we can take any $v_i\rightarrow -v_i$, and so we also have a $(\mathbb{Z}_2)^K$ symmetry. The superpotential $W(v_i)$ must be invariant under both of these symmetries, and so $W$ is a symmetric and even function in $v_1,...,v_K$.

Using this fact, we find that it suffices to prove the existence of the solution $\grf_K$ (for which all of the values on the diagonal are equal, $v_1=...=v_K$) to show that all of the $\grf_n$'s exist. Indeed, since $W$ is even in each of the $v_i$'s, we find that taking any solution and replacing one of the $v_i$'s with zero gives another solution\footnote{The proof of this fact is identical to the proof that an even function $f(x)$ has $f'(0)=0$.}.  Thus it suffices to show that $\grf_K$ exists to show that all of other solutions exist as well.

It should not be surprising that a symmetric function $W$ has such a symmetric extremum $\phi_K$, but we can give a rigorous argument that shows why we expect it to be an extremum to all orders in perturbation theory.

Let us assume that we can show that the solution $\grf_K$ exists at 1-loop (as we have managed to do in the previous sections). We now claim that the form of $\grf_K$ is preserved also to higher loop orders. This is the result of a general lemma, which states that small corrections cannot cause a global symmetry to be broken in the vacuum\footnote{We must assume that the theory is weakly coupled so that perturbation theory is justified, e.g., by taking large $k$. However, since our results remain consistent in the range $k\geq-\min(N,N_f)$, we assume that this works for smaller $k$ as well. It is also important that we have a tree-level mass term for this argument.} \cite{PhysRevD.10.1246}.

We have thus found that the solutions of the form $\grf_n=\begin{pmatrix}
v_n\cdot \id_{n\times n} & \\
& 0
\end{pmatrix}$ always appears, regardless of the specific form of the superpotential. Thus the solutions discussed above remain solutions to all orders in perturbation theory (for some unknown $v_n$'s).

Having proven this, it is simple to show that there is only one phase transition. This is done in Appendix \ref{app_one_phase_transition}. This should already sound reasonable, since the form of the solutions makes it obvious that a phase transition where two vacua $\grf_n,\grf_k$ collide can occur only when $v_n=v_k=0$. Thus the phase transitions all occur at the same point in field space, $\grf=0$ (and in particular, they must include the vacuum at this point).

There are some comments we must make about this proof. First, we can ask whether additional solutions which are not of the form $\phi_n$ can appear at higher loop orders. We do not expect this to happen, since we cannot see these solutions at 1-loop order, and it is very unlikely that small quantum effects can create them. Furthermore, if these vacua exist then they are extremely restricted; for example, we saw that the vacua $\phi_n$ exactly agree with the expected behavior of the Witten index, so that any additional vacuum must have vanishing Witten index, meaning that they can disappear even without a phase transition.

One can also ask why the proof does not work for the large $|m|$ regime as well, where we saw that the only solution is $\grf=0$. Our analysis of the $U(1)$ gauge theory gives us the answer; the solutions $\grf_n$ do appear in this regime, but they become complex solutions (that is, the parameter $v_n$ becomes complex). But then how do we know that these solutions are guaranteed to become real for some range of $m$? This is simply due to the fact that the asymptotic phases do not have the same Witten index. We conclude that the Witten index must jump for some $m$, and so we must have new vacua coming in from infinity. It is reasonable to assume that these vacua are the $\grf_n$'s (precisely because we have shown that their Witten index is the difference between the Witten indices of the two large $|m|$ phases).

Finally, it is important to note that the proof was only valid for matter in the fundamental representation. Indeed, when using adjoint matter one finds completely different solutions \cite{Bashmakov:2018wts}. This is because in the adjoint representation the residual symmetries are different - specifically, after diagonalizing the vev matrix $\grf$, we do not have a residual $(\mathbb{Z}_2)^N$ symmetry.

It is now clear that the proof works for many more gauge groups. In particular, one can redo the argument for gauge groups such as $SO(N),O(N)$ and $Sp(N)$. Indeed, studying the superpotential \eqref{eq:full_superpotential} for these theories, one can explicitly find the solutions $\grf_n$ at 1-loop, and so the proof proceeds in the same manner (in the cases where the superpotential simplifies, as in the $SU(N)\times U(1),SO(N)$ and $O(N)$ cases, one can prove analytically that the $\grf_n$'s are the only solutions in the intermediate phase at 1-loop.). We thus expect similar phase diagrams for these theories as well.

\subsection{Summary}\label{sec:phases_summary}

We summarize the main parts of this section. We have shown explicitly that at 1-loop, both $SU(N)$ and $U(N)$ gauge theories with CS terms and fundamental matter fields all have very similar phase diagrams. The general form of these phase diagram was given in Figure \ref{fig:general_phase_diagram}. 
These phase diagrams have 3 regimes as a function of $m$. For negative mass, there is a single vacuum, found by naively integrating out the massive matter multiplets. For $0<m<m_*$, there are many SUSY vacua, each corresponding to an $\mc{N}=1$ NLSM with a decoupled CS TQFT (these may also be SUSY-breaking). All but one of these vacua appear from infinity in field space for small positive mass, causing the Witten index to jump. At $m=m_*$, these vacua coalesce simultaneously into a single vacuum for $m>m_*$. The vacuum at $m>m_*$ can again be obtained by naively integrating out the massive matter multiplets. The diagram for a $U(N)$ gauge theory is given in Figure \ref{fig:UN_diagram}, and the diagram for an $SU(N)$ gauge theory is given in Figure \ref{fig:SUN_diagram}.
We also considered the special case of an $SU(2)$ gauge theory, where classically the symmetry group is enhanced to $Sp(N_f)$. This changes the symmetry breaking pattern in $SU(2)$ theories as opposed to $SU(N)$ theories with $N>2$. 

It is no accident that all of these theories have similar phase structures, as explained in the symmetry argument in Section \ref{symmetry_argument}. This same symmetry argument proves that all of the different vacua in the intermediate phase all coalesce simultaneously at $m=m_*$ to all orders in perturbation theory (this is highly unnatural, and we understand this better after studying the $\mathcal{N}=2$ versions of these theories in the next sections). We thus expect the 1-loop picture to persist to all orders in perturbation theory, and expect quantum effects to change this picture only slightly (e.g. by shifting the value of $m_*$ from the 1-loop result).
The fact that there is a single phase transition also allows us to suggest dual descriptions for the phase transition at $m=m_*$. We take a more careful look at this point in the following sections. 

\section{Fixed Points and RG Flows at Weak Coupling}
\label{sec:RG_flows}

In the previous sections we analyzed $\mc{N}=1$ $SQCD_3$ with a mass deformation, and found a second order phase transition (which should be described by a CFT). We now attempt to study this fixed point more carefully, in the limit where the theory is weakly coupled (in particular, for large CS level $k$). In this limit we can safely decouple the Yang-Mills term, and are left with a pure CS-matter theory. We can then use the results of  \cite{AVDEEV1992561,avdeev1993renormalizations}, where the beta function of a CS theory with matter fields was analyzed.  

We start by considering the general RG flow diagram obtained in \cite{avdeev1993renormalizations}, and determine the specific fixed point our theory flows to at large $k$, assuming the classical superpotential vanishes in the CS-matter theory. In particular, we find that for some degenerate cases, this fixed point has emergent $\mc{N}=2$ SUSY. We then consider the special case of an $SU(2)$ gauge theory. We find that the enhanced symmetry observed in Section \ref{su(2) phase} is also apparent in the RG flow diagram, and supports our previous analysis. 

There is one important subtlety we must take care of before applying the results of \cite{avdeev1993renormalizations} to $\mc{N}=1$ $SQCD_3$. When we decouple the Yang-Mills term and keep only the CS term  from $\mc{N}=1$ $SQCD_3$ (see Appendix \ref{appendix YM and CS}), we may generate a superpotential for the resulting CS theory, and in particular, generate non-zero values for the marginal couplings. Thus, to find the correct RG fixed point we flow to, we must plug in these couplings into the result of \cite{avdeev1993renormalizations}, and follow the RG flow to find the fixed point. However, we will show that at large $k$, the couplings which are generated are negligible, and so our assumption that the flow starts with vanishing classical superpotential in the CS-matter theory is justified.

\subsection{RG Flows and Fixed Points at Large \texorpdfstring{$k$}{k}}
\label{Rg su(n)}

We start by analyzing the fixed point that our $\mc{N}=1$ $SQCD_3$ theory flows to at large $k$ (the specific case of an $SU(2)$ gauge group is studied in the next section). We emphasize once again that we assume that in flowing from the YM-CS theory to the CS theory, the additional couplings that are generated may be neglected (we prove that this assumption is justified in Section \ref{initial point}). We can thus find the fixed point we flow to by following the RG flows of \cite{avdeev1993renormalizations} when starting from the origin.

There are two cases we must discuss separately. For $\mc{N}=1$ $SQCD_3$ with gauge group $SU(N)$ or $U(N)$, there are generically two classically marginal operators which preserve $\mc{N}=1$ SUSY; these are $\frac{1}{4}\eta_0(\bar\Phi^i\Phi_i)^2$ and $\frac{1}{4}\eta_1(\bar\Phi^i T^a \Phi_i)^2$, where $i=1,...,N_f$ is a flavor index and $\eta_0,\eta_1$ are coupling constants. However, in degenerate cases, these two operators coincide. These degenerate cases are:
\begin{itemize}
\item When the gauge group is $U(1)$
\item When $N_f=1$
\end{itemize}
In these two cases, there is only one marginal operator that preserves $\mc{N}=1$. So we must discuss the degenerate and the non-degenerate cases separately.

We start by discussing the degenerate cases. The results of \cite{AVDEEV1992561,avdeev1993renormalizations} show that in these cases, flows that originate in a neighborhood of the origin flow to a fixed point with $\mc{N}=2$ SUSY, which is located at $\eta_1=\frac{8\pi}{k}$. These theories thus have emergent $\mc{N}=2$ SUSY at the fixed point.

\begin{figure}[h]
     \subfloat[\label{subfig:RGflow_a}]{%
       \includegraphics[width=8.5cm, height=5cm]{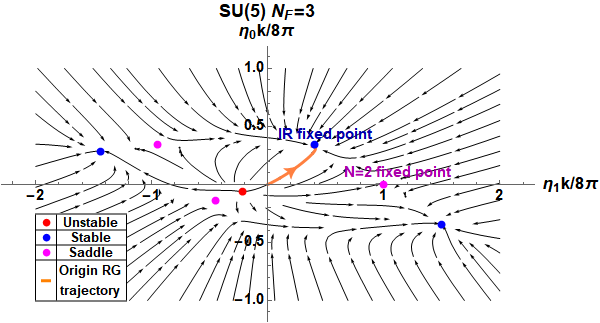}
     }
     \hspace{5mm}
     \subfloat[\label{subfig:RGflow_b}]{%
       \includegraphics[width=8.5cm, height=5cm]{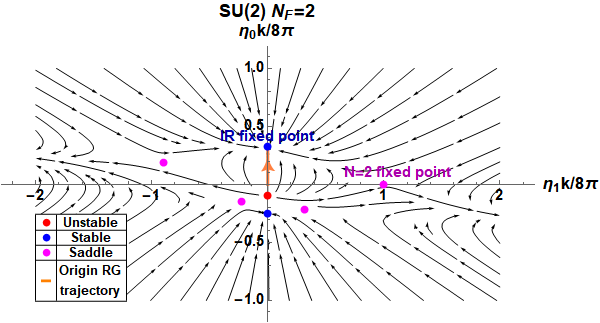}
     }
     \caption{\small RG flows for (a) $SU(5)$ with $N_f=3$ and (b) $SU(2)$ with $N_f=2$. These represent the general structure of the RG flows at large $k$. The $\mathcal{N}=2$ fixed point is at $(1,0)$, which is a saddle point for $N_f\geq 2$. For $N>2$ there is a stable $\mathcal{N}=1$ fixed point in the first quadrant, which attracts flows starting near the origin. For $N=2$, a stable $\mathcal{N}=1$ fixed point along the positive vertical axis attracts flows originating near $(0,0)$.}
     \label{fig:RGflow}
\end{figure}

We now discuss the non-degenerate cases, where there are two independent classically marginal operators. For matter fields in the fundamental representation, the results of  \cite{avdeev1993renormalizations} give the following result for the beta function for an $SU(N)$ gauge theory with $N_f>1$  (using $\tilde{k}\equiv \frac{k}{2\pi}$):
\begin{eqnarray} \label{beta function}
\beta_{\eta_1}^{SU(N), N_f}&\!\!=\!\!&\frac{1}{8 N^2 \tilde{k}^3}\left\{ \eta_1^3 \tilde{k}^3 \left((3 N^2-10) N N_f-5N^2+22\right)\right.
\nonumber\\[-1mm]
&&\qquad\qquad+8 \eta_1 \tilde{k} \left[N \left(\eta_0^2 N
   \tilde{k}^2 \left(2 N N_f+11\right)+4 \eta_0
   \left(N^2-1\right) \tilde{k}-6 \left(N^2-2\right)
   N_f\right)+10 N^2-28\right]
\nonumber \\
&&\left.\qquad\qquad -64(N^2-4)(\eta
  _0 N \tilde{k}+N N_f-3)+4 \eta_1^2 \tilde{k}^2
   \left(\eta_0 N \tilde{k}(3 N N_f+7 N^2-22)+N^3
   N_f-3 N^2+4\right)\right\}
\nonumber\\[3mm]
\beta_{\eta_0}^{SU(N), N_f}&\!\!=\!\!&\frac{1}{16 N^3 \tilde{k}^3} \left\{\eta_1^3 (N^2-1) \tilde{k}^3 (2 N N_f+3
   N^2-10)+4 \eta_1^2 (N^2-1) \tilde{k}^2
   \left(2 \eta_0 N \tilde{k} (NN_f+2)+N^2-2\right)
 \right.  \nonumber\\
   &&\qquad\qquad+16 \left[3 \eta_0^3 N^3
   \tilde{k}^3 (N N_f+2)-8 \eta_0 N^2
   (N^2-1) \tilde{k} N_f+2 \eta_0^2 N^2
   (N^2-1) \tilde{k}^2\right.
  \nonumber \\
   &&\left.\left.\qquad\qquad\qquad-4 (N^2-1)(2
   N N_f+N^2-6)\right]+8 \eta_1 \left(N^2-1\right)
   \tilde{k} \left(7 \eta_0^2 N^2 \tilde{k}^2-6
   N^2+12\right)\right\}
\end{eqnarray}
the results for a $U(N)$ gauge theory are similar in the large $k$ limit. We plot the resulting RG flow in Figure \ref{subfig:RGflow_a}.

We now discuss the fixed points. First, we find that in the non-degenerate case, the $\mathcal{N}=2$ fixed point is actually a saddle point, and so is unstable. Instead, there is a stable $\mathcal{N}=1$ fixed point located at $\eta_0,\eta_1>0$. From inspection of the beta function or Figure \ref{subfig:RGflow_a}, we can see that flows which originate in a small neighborhood of the origin end up at this $\mathcal{N}=1$ fixed point. We conclude that the non-degenerate cases of $\mc{N}=1$ $SQCD_3$ flow to this $\mathcal{N}=1$ fixed point. 

To summarize, we studied the fixed point for $\mc{N}=1$ $SQCD_3$ with large CS level $k$ and no classical superpotential. In particular, we find that for large enough $k$:
\begin{itemize}
\item A $U(1)_k$ gauge theory with $N_f\geq 1$ flows to an $\mathcal{N}=2$ fixed point.
\item $SU(N)_k$ and $U(N)_k$ gauge theories with $N>1$ and $N_f= 1$ flow to an $\mathcal{N}=2$ fixed point.
\item $SU(N)_k$ and $U(N)_k$ gauge theories with $N>1$ and $N_f> 1$ flow to an $\mathcal{N}=1$ fixed point.
\end{itemize}

\subsection{RG Flows for \texorpdfstring{$SU(2)$}{SU2} and Enhanced Global Symmetry}
\label{RG su(2)}

For an $SU(2)$ gauge theory, the beta function simplifies (here again, $\tilde{k}\equiv\frac{k}{2\pi}$):
\begin{eqnarray}
\label{beta function 2}
\beta_{\eta_1}(\eta_1,\eta_0)\!&\!\!=\!\!&\!
\frac{\eta_1 \tilde{k}}{16 \tilde{k}^3}
\left\{ \left[\eta_1^2 \tilde{k}^2 (2
   N_f+1)+8 \eta_1 \tilde{k} \left(3 \eta_0 \tilde{k}
   (N_f+1)+2 (N_f-1)\right]\right.\right.
  \nonumber \\
   &&\left.\left.\qquad~~ +16 \left[\eta
  _0^2 \tilde{k}^2 (4 N_f+11)+6 \eta_0 \tilde{k}-6
   N_f+3\right)\right]\right\}\nonumber\\[3mm]
\beta_{\eta_0}(\eta_1,\eta_0)\!&\!\!=\!\!&\!\frac{3}{64 \tilde{k}^3}\left\{ \eta_1^3 \tilde{k}^3 (2 N_f+1)
+4 \eta_1^2 \tilde{k}^2 \left(4 \eta_0 \tilde{k}
   (N_f+1)+1\right)\right.\nonumber\\
  &&\left.\qquad~~+64 \left[2 \eta_0^3 \tilde{k}^3
   (N_f+1)-4 \eta_0 \tilde{k} N_f+\eta_0^2
   \tilde{k}^2-2 N_f+1\right]
 +16 \eta_1 \tilde{k} (7 \eta
  _0^2 \tilde{k}^2-3)\right\}
\end{eqnarray}
The corresponding RG flow is shown in Figure \ref{subfig:RGflow_b}. We find that the $\mathcal{N}=2$ fixed point is still unstable, but now the stable $\mathcal{N}=1$ fixed point lies on the vertical axis ($\eta_0=0$). 
Furthermore, we find $\beta_{\eta_1}(0,\eta_0)=0$, which means that if the theory has $\eta_1=0$ classically then no $\eta_1$ is generated by quantum effects. This is consistent with the fact that $\eta_0\neq0$ preserves the $Sp(N_f)$ global symmetry discussed in Section \ref{su(2) phase}, while $\eta_1\neq0$ does not. This gives a non-trivial check for the validity of the calculation in \cite{avdeev1993renormalizations}.

We conclude that an $SU(2)$ gauge theory with $N_f>1$ at large $k$ flows to an $\mathcal{N}=1$ fixed point which preserves the $Sp(N_f)$ global symmetry.

\subsection{Flowing from Yang-Mills-Chern-Simons to Chern-Simons Theory at Large \texorpdfstring{$k$}{kk}}
\label{initial point}

We now justify our assumption that the correct fixed point for our theory is found by setting the classical superpotential of the corresponding CS-matter theory to zero. As explained above, our theories differ from those studied in \cite{avdeev1993renormalizations} by a YM term. To find the correct fixed point for our theories, we must find what couplings are generated by integrating out the YM term. We find that these couplings are negligible for large $k$. This means that it was justified to find the fixed point by starting the flow from the origin as done in Figure \ref{fig:RGflow}.

In flowing from YM-CS theory to CS theory, we generate $\grF^4$ couplings in the effective action\footnote{We also generate some non-renormalizable terms, which we ignore, as well as a shift of the CS level of order $N$, which we can ignore at large $k$.}. To find the couplings, we must find the effective action which results from integrating out only the very massive gauge field mode. To explain how this is done in more detail, we assume that the gauge group is $U(1)$ for simplicity. At large $k$ an abelian Higgsed YM-CS theory has two massive gauge field modes, with masses \cite{Dunne:1998qy}: 
$$ m_\pm=\frac{m_{CS}}{2}\left(\sqrt{1+\frac{4m_H^2}{m_{CS}^2}}\pm1\right) $$
Here, $m_{CS}=\frac{kg^2}{2\pi}$ and $m_H=2gv$ with $v$ the scalar vev.
We find that at large CS level $k$, we have one very massive mode (with mass $m\sim kg^2$) and one light mode. Comparing to the spectrum of a Higgsed CS theory, we find that we can associate the very massive mode with the YM term and the light mode with the CS term. Since we want to flow to a CS theory, we must integrate out only the very massive YM mode to end up with a Higgsed CS theory. We show how this is achieved in Appendix \ref{appendix YM and CS} (the same method applies for a nonabelian gauge group as well). 

Integrating out only the very massive mode, we generate nonzero $\eta_0$  and $\eta_1$ (call these values $=\eta_0^{(0)},\eta_1^{(0)}$ respectively). Furthermore, we find $\eta_i^{(0)}\propto\frac{1}{k^2}$ for $i=1,2$. One can use a dimensional analysis argument to prove this. First, note that in the effective action for $\Phi$ when integrating out only the gauge field, each $\Phi$ external leg comes with a factor of $g$. Since the only other dimensionful parameters come from the combination $\kappa=\frac{kg^2}{2\pi}$, the only way to give the effective action the correct dimensions is to have a factor of $\kappa^2$ in the denominator. We thus find $W|_{\Phi^4}\propto \frac{(g\Phi)^4}{\kappa^2}$, which leads to the required result. 

We now show that since $\eta_i^{(0)}\propto\frac{1}{k^2}$, we must flow to the attractive fixed point in the first quadrant of Figure \ref{fig:RGflow}.
Consider the beta functions \eqref{beta function},\eqref{beta function 2}. These can be put in the form $\beta_{\eta_{1,2}}=\frac{1}{k^3}f_{1,2}(\eta_1k,\eta_2k)$ for some functions $f_1,f_2$. We conclude that the RG flow only depends on $\eta_1,\eta_2$ through the combinations $\eta_i\cdot k$, and so in particular the fixed points appear at $\eta_i\sim\frac1k$. Since our starting point is at $\eta_i^{(0)}\propto\frac{1}{k^2}$, we can conclude that by taking $k$ to be large enough, we can set the starting point of the RG flow in Figure \ref{fig:RGflow} to be arbitrarily close to the origin. Since there exists a neighborhood of the origin which flows to the same fixed point, we find that at large enough $k$ we must flow to this fixed point. 

To summarize, we have managed to relate our YM-CS theory with matter to the CS-matter theory studied in \cite{avdeev1993renormalizations}. Specifically, even though integrating out the YM term generates an effective superpotential, we have found that at large $k$ we can neglect this superpotential when looking for the starting point of the RG flow. We thus find that our assumption that we can flow from the origin to find the correct fixed point is justified.

\section{Dualities}\label{duality_for_general_Nf}

\setcounter{equation}{0}
Using the phase diagrams described above, one can generalize the $N_f=1$ duality described in \cite{Bashmakov:2018wts} to larger $N_f$. The generalization is:
\begin{equation}\label{duality}
U(N)_{k+\frac12 N+\frac12 N_f,k+\frac12 N_f}+N_f\;\Phi \longleftrightarrow SU(k+N_f)_{-N-\frac12 k}+N_f\;\widetilde{\grF}
\end{equation}
With $\grF\grF$ matching to $-\widetilde{\grF}\widetilde{\grF}$. Here we shall provide evidence for this duality for all $N_f\geq 1$ and $k>-\min(N_f,N)$.

Let us sketch a general picture of the phases of both sides of the duality, consistent with the 1-loop superpotential analysis done above. We have shown that at 1-loop, there is a single phase transition point located at $m=m_*$. On one side of the transition, both sides of the duality \eqref{duality} have a single vacuum: on the LHS this is a $U(N)_{k+N_f}$ TQFT, and on the RHS this is an $SU(k+N_f)_{-N}$ TQFT (schematically, this is the rightmost phase in Figure \ref{fig:general_phase_diagram}). These are level rank dual, and so they match.

On the other side of the transition, the picture is much more complicated. Both sides of the duality have many vacua, which collide simultaneously at $m=m_*$ (this is the intermediate phase in Figure \ref{fig:general_phase_diagram}). Due to this simultaneous collision (or equivalently, since there is only phase transition), all of these vacua must appear arbitrarily close to the phase transition point, and so it seems like we would need to match all of these vacua across the duality. However, we remember that some of these vacua can be SUSY-breaking, and so they do not participate in the phase transition\footnote{Even if these SUSY-breaking vacua aren't removed by higher-order effects, their energy is necessarily larger than that of the SUSY-preserving vacua, meaning that they cannot take part in this second-order phase transition.}. 
Indeed, we are able to match the SUSY-preserving vacua on both sides of the duality.

In the following sections, we study the phases of the two sides of the duality, and show that we can match all of the SUSY-preserving phases across the two sides. We note that the matching of the phases here is a direct extension of the matching of the phases in the case $N_f=1$ discussed in Section \ref{review_of_adjoint_paper}. In the $N_f=1$ case, a Higgsed vacuum was matched with a non-Higgsed vacuum. We find a similar result here; a $ U(N) $ vacuum which is Higgsed to $ U(N-n) $ corresponds to an $ SU(N) $ vacuum which is Higgsed to $ SU(N-(N_f-n)) $ (where we have suppressed the CS levels). We also emphasize that for the phases to match, it is imperative that there is only one phase transition, so that all of the vacua coalesce at the same point on both sides.

Finally, we show that most our $\mathcal{N}=1$ dualities can be understood as resulting from some known $\mathcal{N}=2$ dualities. This might not be too much of a surprise, since we saw that the $\mc{N}=1$ fixed point is close to the $\mc{N}=2$ fixed point in the RG sense, and so we might expect that there exists an RG flow from the $\mc{N}=2$ point to the $\mc{N}=1$ point which could explicitly prove the relation between the dualities. We also find that one of the dualities does not have an $\mc{N}=2$ version in the literature. We describe this $\mc{N}=2$ duality and discuss some nontrivial checks we have performed on it.

\subsection{Warmup: \texorpdfstring{$N=1$}{n1}}

In the case $N=1$, the $\mc{N}=1$ duality \eqref{duality} becomes
\beq
U(1)_{k+\frac12 N_f}+N_f\;\Phi \longleftrightarrow SU(k+N_f)_{-\frac12 k-1}+N_f\;\widetilde{\grF}
\eeq{}
Let us match the phases for $k>0$. As we saw above, a $U(1)$ gauge theory has a phase transition at some $m=m_*$. On one side of this transition we have two vacua ($U(1)_{k}$ and a $\mathbb{CP}^{N_f-1}$ NLSM), and on the other we have a single vacuum ($U(1)_{k+N_f}$).
As for the $SU(k+N_f)$ gauge theory, we also have a phase transition at some $m=m_*$. On one side of this transition we have a single vacuum ($SU(k+N_f)_{-1}$), while on the other side we have $N_f+1$ vacua. However, only two of these vacua are SUSY-preserving: $SU(k)_{-1}$ and $\mathcal{M}_{N_f,1}=\mathbb{CP}^{N_f-1}$. 
Using level-rank duality, we can thus exactly match the phases of these two theories around the transition point.

We are left with matching the phases for\footnote{the case $N=1,N_f=1$ is more subtle and is discussed in \cite{Bashmakov:2018wts}. Here we discuss $N_f>1$} $k=0$. While the phase with a single vacuum is unchanged and can still be matched across the duality, we find that the vacua in the intermediate phase are slightly modified. On the $U(1)$ side, the vacua are now a $U(1)_0$ gauge theory and a $\mathbb{CP}^{N_f-1}$ NLSM. On the $SU(N)$ side, the vacua are now a $\mathbb{CP}^{N_f-1}$ and a $ \frac{U(N_f)}{SU(N_f)}=S^1$ NLSM. We can immediately match the two $\mathbb{CP}^{N_f-1}$ vacua. To match the remaining vacua, we notice that $U(1)_0$ has a circle of vacua, parametrized by the dual scalar. We can thus match the two remaining vacua, since they are both NLSMs with target space $S^1$.

\subsection{Matching the Phases for general \texorpdfstring{$N$}{n}}

We now match the phases across the phase transition point for general $N$. Referring to Figure \ref{fig:general_phase_diagram}, we find that both theories have one vacuum on one side of the transition point, and many on the other side. We have already discussed the matching for the side with the single vacuum, and we are left with matching the many vacua in the intermediate phase across the duality. There are four regimes we must do this for:

\begin{itemize}
\item $\pmb{N_f<N}$ \textbf{and} $\pmb{k>0}$. This is the simplest form of the duality. Referring to Sections \ref{phases_of_SUN} and \ref{phases_of_UN}, we find that all of the vacua are SUSY-preserving. The vacua are matched as
\begin{equation} \label{eq: Vector Multiplet Duality} 
\mathcal{M}_{N_f,n} \times U(N-n)_{k+\frac12 (N+n),k+n}^{\mc{N}=1}
\longleftrightarrow 
\mathcal{M}_{N_f,N_f-n} \times SU(k+n)_{-N+\frac12(n-k)}^{\mc{N}=1},
~~~~n= 0,...,N_f
\end{equation}
Indeed, since $k>0$, this is just the $\mc{N}=1$ version of level-rank duality described in equation \eqref{eq:N1_level_rank} (note that by definition, $\mathcal{M}_{N_f,N_f-n}=\mathcal{M}_{N_f,n}$).

\item $\pmb{N_f<N}$ \textbf{and} $\pmb{-N_f<k\leq 0}$.
Referring to the phases described above, we find the following phase structure. On the $U$ side we find $|k|$ SUSY-breaking vacua and $N_f-|k|+1$ SUSY-preserving vacua of the form $\mathcal{M}_{N_f,n} \times U(N-n)_{k+\frac12 (N+n),k+n}^{\mc{N}=1}$ with  $n=|k|,...,N_f$. We note that the vacuum with $n=|k|$ is $\mathcal{M}_{N_f,|k|} \times U(N+k)_{\frac12 (N+k),0}^{\mc{N}=1}$, and so its index vanishes.

On the $SU$ side, we find $N_f-|k|$ SUSY-preserving vacua of the form $\mathcal{M}_{N_f,N_f-n} \times SU(k+n)_{-N+\frac12 (n-k)}^{\mc{N}=1}$ with $n=|k|+1,...,N_f$, and one SUSY vacuum $\frac{U(N_f)}{SU(N_f-|k|)\times U(|k|)}$ with vanishing index.

We can use the $\mc{N}=1$ version of level-rank duality to match all of the phases, except for the phase with vanishing Witten index which needs some care\footnote{\label{footnote_match}Using the fact that a $U(1)_0$ gauge theory is dual to a free scalar, we find that the target spaces of the two phases are locally both $\frac{U(N_f)}{SU(N_f-|k|)\times U(|k|)}$. However, there are some global effects one must be careful with. We assume that a more careful analysis of these phases matches the global behavior of the target spaces as well.}.

\item $\pmb{N_f\geq N}$ \textbf{and} $\pmb{k>0}$. We can once again refer to the phases described above. On the $U$ side we have $N+1$ SUSY-preserving vacua of the form $ \mathcal{M}_{N_f,n} \times U(N-n)_{k+\frac12 (N+n),k+n}^{\mc{N}=1}$. On the $SU$ side we find $N_f-N$ SUSY-breaking vacua, and $N+1$ SUSY-preserving vacua of the form  $\mc{M}_{N_f,N_f-n}\times
SU(k+n)_{-N+\frac12 (n-k)}^{\mc{N}=1}$. Again, using the $\mc{N}=1$ version of level-rank duality \eqref{eq:N1_level_rank} we find that these phases match.

\item {$\pmb{N_f\geq N}$ \textbf{and} $\pmb{-N<k\leq 0}$}.
On the $U$ side we find $|k|$ SUSY-breaking vacua and $N-|k|+1$ SUSY-preserving vacua of the form $ \mathcal{M}_{N_f,n} \times U(N-n)_{k+\frac12(N+n),k+n}^{\mc{N}=1}$ for $n=|k|,...,N$. Again, we find that for $n=|k|$ we must take special care of the SUSY vacuum $\mathcal{M}_{N_f,|k|} \times U(N+k)_{\frac12(N+k),0}^{\mc{N}=1}$, since its Witten index vanishes.

On the $SU$ side we find $N_f-N$ SUSY-breaking vacua,
$N-|k|$ SUSY-preserving vacua of the form  $\mc{M}_{N_f,N_f-n}\times
SU(k+n)_{-N+\frac12(n-k)}^{\mc{N}=1}$, and a single SUSY-preserving vacuum of the form
$\frac{U(N_f)}{SU(N_f-|k|)\times U(|k|)}$
(with vanishing index).
We once again find that apart from using the $\mc{N}=1$ level-rank duality to show that the phases match, we must also take special care of the vacua with vanishing Witten index\textsuperscript{\ref{footnote_match}}.

\end{itemize}

Finally, we comment on the case $N_f>N$ and $k<-N$. In this case, we find some SUSY-preserving vacua for which we cannot use the known level-rank duality to match the phases.

\subsection{More Dualities}\label{more_dualities}

We can now generalize this discussion to other gauge groups. Due to the symmetry argument in Section \ref{symmetry_argument}, we expect $SO(N)$ and $Sp(N)$ gauge groups to have the same general form for solutions to the F-term equations (indeed, a numerical analysis confirms this). We can thus find the phase diagrams for these theories. Repeating the analysis above for these gauge groups leads to the following $\mc{N}=1$ dualities:
\begin{align}
SO(N)_{k+\frac12(N-2+N_f)}+N_f\;\grF&\longleftrightarrow SO(k+N_f)_{-\frac12(k-2)-N}+N_f\;\widetilde{\grF}\label{eq:SON_duality}\\
Sp(N)_{k+\frac12(N+1+N_f)}+N_f\;\grF&\longleftrightarrow Sp(k+N_f)_{-\frac12(k+1)-N}+N_f\;\widetilde{\grF}\label{eq:SpN_duality}
\end{align}
We will not give a detailed proof here that the phases match for the various ranges of parameters, since the details are very similar to those discussed above. Instead, we restrict ourselves to the simplest case of $k\geq 0$ and $N_f<N$. In this case, there are no SUSY-breaking phases. The phases for the theories in the $SO(N)$ duality \eqref{eq:SON_duality} with $k\geq 0,\;N_f<N$ are summarized in the following table\footnote{We are being purposefully careless here with the definition of $m$ so as to simplify notation. Specifically, we are being careless with the sign of $m$. Note that the duality matches $m\Rightarrow-m$.}:
\begin{center}
\begin{tabular}{c c c}
& LHS & RHS \\ 
\hline
$m>m_*$& $SO(N)_{k+N_f}$&$SO(k+N_f)_{-N}$ \\ 
$m<m_*$& $SO(N-n)_{k+n}\times{\widetilde{\mc{M}}_{N_f,n}}$ & $SO(k+N_f-n)_{-N+N_f-n}\times{\widetilde{\mc{M}}_{N_f,n}}$
\end{tabular}
\end{center}
where $n=0,...,N_f$ and we used $\widetilde{\mc{M}}_{N_f,n}=\frac{SO(N_f)}{S[O(n)\times O(N_f-n)]}$ as shorthand for a NLSM with target space $\widetilde{\mc{M}}_{N_f,n}$. Just like in the $SU(N)$ and $U(N)$ cases discussed above, we find that using level-rank duality we can match the $n$'th vacuum on the LHS with the ($N_f-n$)'th vacuum on the RHS (one can also compare to \cite{Aharony:2016jvv}).

We can also repeat the discussion above to find dualities between $U(N)$ groups. Indeed, using $U$-$U$ level-rank dualities \cite{Hsin:2016blu}, a similar procedure leads to the $\mc{N}=1$ dualities 
\begin{align}
U(N)_{k+\frac12(N+N_f),~k+N+\frac12 N_f}+N_f\;\grF &\longleftrightarrow U(k+N_f)_{-N-\frac12 k,~-N-k-\frac12 N_f}+N_f\;\widetilde{\grF}\\
U(N)_{k+\frac12(N+N_f),~k-N+\frac12 N_f}+N_f\;\grF &\longleftrightarrow U(k+N_f)_{-N-\frac12 k,~k-N+\frac{3}{2} N_f}+N_f\;\widetilde{\grF}
\end{align}
Once again, we only show the matching of the phases for the simplest case $k\geq0,\;N_f<N$. In this case the phases for the two sides of the first duality are:
\begin{center}
\begin{tabular}{c c c}
& LHS & RHS \\ 
\hline
$m>m_*$& $U(N)_{k+N_f,~k+N+N_f}$&$U(k+N_f)_{-N,~-k-N-N_f}$ \\ 
$m<m_*$& $U(N-n)_{k+n,~k+N}\times\mc{M}_{N_f,n}$ & $U(k+N_f-n)_{-N+N_f-n,~-k-N}\times\mc{M}_{N_f,n}$
\end{tabular}
\end{center}
and for the second duality:
\begin{center}
\begin{tabular}{c c c}
& LHS & RHS \\ 
\hline
$m>m_*$& $U(N)_{k+N_f,~k-N+N_f}$&$U(k+N_f)_{-N,~k-N+N_f}$ \\ 
$m<m_*$& $U(N-n)_{k+n,~k-N+2n}\times\mc{M}_{N_f,n}$ & $U(k+N_f-n)_{-N+N_f-n,~k-N+2N_f-2n}\times\mc{M}_{N_f,n}$
\end{tabular}
\end{center}
where $n=0,...,N_f$. Once again, using level-rank dualities we can match the $n$'th vacuum on the LHS with the ($N_f-n$)'th vacuum on the RHS.

\subsection{Emergent Symmetries and Supersymmetry}

We now comment on some interesting consequences of our duality $\eqref{intro_duality}$, which are related to emergent global symmetries, supersymmetry, and time-reversal symmetry in the IR. Similar results were found within the context of $\mathcal{N}=1$ SUSY in \cite{Gaiotto:2018yjh, benini2018n}. Finally we analyze the special cases of $U(1)\simeq SO(2)$ and $SU(2)\simeq Sp(1)$ to obtain a series of dualities similar to the non-SUSY case in \cite{Benini:2017dus}.

Our conjectured duality \eqref{intro_duality} is a strong-weak duality\footnote{This can be easily seen when we change regularization schemes; switching from Yang-Mills regularization of the CS level $k$ to dimensional regularization with CS level $\kappa$ following \cite{Aharony:2015mjs}, the 't Hooft coupling $\lambda=\frac{N}{k}$ of the two sides of the duality are related by $|\lambda_{LHS}|=1-|\lambda_{RHS}|$.}. 
Thus, we cannot use the RG flow analysis above to describe both sides of the duality (since the analysis assumes large $k$). However, assuming the duality is correct, we can find some interesting conclusions about the strongly coupled side by analyzing the weakly coupled side.

We start by discussing emergent global symmetries. As discussed in \ref{su(2) phase}, an $SU(2)$ gauge theory with $N_f$ fundamentals has classically enhanced $Sp(N_f)$ flavor symmetry. Similarly, an $SO(2)$ gauge theory with $N_f$ fundamentals has enhanced $U(N_f)$ symmetry.
Combined with the $SU$-$U$ and $SO$-$SO$ dualities \eqref{intro_duality},\eqref{intro_SO_duality} we find the following emergent symmetries:
\begin{align}
U(N)_{2+\frac12(N-N_f),2-\frac12 N_f}&\longrightarrow~ \text{emergent }Sp(N_f)\text{ global symmetry}
\\SO(N)_{1+\frac12(N-N_f)}&\longrightarrow~ \text{emergent } U(N_f) \text{ global symmetry}
\end{align}
in the range where the dualities are valid.

Next we discuss emergent supersymmetry.
In Section \ref{sec:RG_flows} we saw that in the weak coupling regime with large CS level $k$, the theories with $N_f=1$ and the theories with $U(1)\simeq SO(2)$ gauge symmetry flow to an $\mathcal{N}=2$ IR fixed point.
Combined with the dualities above, we can conjecture that in the range where our dualities are correct (and also for large enough $N$), we have:
\begin{equation}
\begin{aligned}
&SU(N)_{1+\frac12(N-N_f)}+N_f ~\Phi ~\longrightarrow ~
 \text{emergent }\mathcal{N}=2\text{  SUSY}
\\
&SO(N)_{1+\frac12(N-N_f)}+N_f ~\Phi ~\longrightarrow ~
\text{emergent }\mathcal{N}=2\text{ SUSY}
\end{aligned}
\end{equation}

Next, we discuss emergent time-reversal symmetry.
Using the dualities of \eqref{intro_duality}-\eqref{intro_U_duality_2} with $k=N-N_f$, we find the following dualities:
\beq
\begin{array}{rcl}
SO(N)_{\frac32N-\frac12N_f-1}+N_f\;\grF & \longleftrightarrow & SO(N)_{-\frac32N+\frac12N_f+1}+N_f\;\widetilde{\grF}\\
Sp(N)_{\frac32 N-\frac12 N_f+\frac12}+N_f\;\grF & \longleftrightarrow & Sp(N)_{-\frac32 N+\frac12 N_f-\frac12}+N_f\;\widetilde{\grF}\\
U(N)_{\frac32N-\frac12N_f,~2N-\frac12N_f}+N_f\;\grF &\longleftrightarrow & U(N)_{-\frac32N+\frac12N_f,~-2N+\frac12N_f}+N_f\;\widetilde{\grF}\\
U(N)_{\frac32N-\frac12N_f,~-\frac12N_f}+N_f\;\grF &\longleftrightarrow & U(N)_{-\frac32N+\frac12N_f,~+\frac12N_f}+N_f\;\widetilde{\grF}
\end{array}
\eeq{dualities}
Note that in each duality, the two sides are related by time reversal. We thus conclude that these theories have emergent time reversal symmetry in the IR. Due to the fact that the superpotential of 2+1$d$ $\mathcal{N}=1$ theories must be time-reversal odd, these theories might have exact moduli spaces \cite{Gaiotto:2018yjh}. Indeed, a special case of the third duality of \eqref{dualities} is $N=N_f=1$, which tells us that the theory $U(1)_{3/2}+\grF$ has emergent time-reversal symmetry. This theory was studied in \cite{Gaiotto:2018yjh}, and was shown to be dual to a $U(1)_0$ gauge theory with a charge 2 matter field, so that it is indeed time-reversal invariant. As a result, it was also shown that this theory has an exact quantum moduli space. It might be possible to find similar dualities that relate the theories in \eqref{dualities} to theories that are manifestly time-reversal invariant. Consequently, a similar analysis might prove that these theories all have exact quantum moduli spaces.

Finally, we discuss a simple series of dualities.
Using $U(1)\simeq SO(2)$ and $SU(2)\simeq Sp(1)$, we can obtain the following two series of dualities using \eqref{intro_duality}$-$\eqref{intro_U_duality_2}. Note that the range of validity of the $SU$-$U$ duality forces us to have $N_f\leq 2$ (a similar non-SUSY version was discussed in \cite{Benini:2017dus})
\begin{alignat*}{6}
U(3)_{-2,\frac52}+\Phi &\;\;\longleftrightarrow\;\;& U(3)_{2,-\frac52}+\Phi &\;\;\longleftrightarrow\;\;& U(1)_{-\frac32}+\Phi &\;\;\longleftrightarrow\;\;& U(1)_{\frac{3}{2}}+\Phi &\;\;\longleftrightarrow\;\;&  SU(2)_{-\frac32}+\Phi &\;\;\longleftrightarrow\;\;& SU(2)_{\frac{3}{2}}+\Phi\\
U(3)_{-\frac{3}{2},3}+2\Phi &\;\;\longleftrightarrow\;\;& U(3)_{\frac{3}{2},-3}+2\Phi &\;\;\longleftrightarrow\;\;& U(1)_{-1}+2\Phi &\;\;\longleftrightarrow\;\;& U(1)_1+2\Phi &\;\;\longleftrightarrow\;&  SU(2)_{-1}+2\Phi &\;\;\longleftrightarrow\;\;& SU(2)_1+2\Phi
\end{alignat*}
In particular we note that both series have emergent time reversal symmetry in the IR. Furthermore, the first series has emergent $Sp(1)$ flavor symmetry, while the second one has emergent $Sp(2)$ flavor symmetry (this becomes apparent when studying the theories in the series with $SU(2)$ gauge groups).

\subsection{Relation to \texorpdfstring{$\mc{N}=2$}{n=2} Dualities}

We now discuss the relation between the $\mc{N}=1$ dualities discussed above and some well-known $\mc{N}=2$ dualities. We focus on the $\mc{N}=1$ dualities for $SU(N)$ and $U(N)$ gauge groups (the results for $SO$ and $Sp$ gauge groups are similar). These are:
\begin{align}
U(N)_{k+\frac12(N+N_f),~k+\frac12 N_f}+N_f\;\Phi &\longleftrightarrow SU(k+N_f)_{-N-\frac12 k}+N_f\;\widetilde{\Phi}\label{eq:U_SU_duality}\\
U(N)_{k+\frac12(N+N_f),~k+N+\frac12 N_f}+N_f\;\grF &\longleftrightarrow U(k+N_f)_{-N-\frac12 k,~-N-k-\frac12 N_f}+N_f\;\widetilde{\grF}\label{eq:first_U_U_duality}\\
U(N)_{k+\frac12(N+N_f),~k-N+\frac12 N_f}+N_f\;\Phi &\longleftrightarrow U(k+N_f)_{-N-\frac12 k,~k-N+\frac32 N_f}+N_f\;\widetilde{\Phi}\label{eq:second_U_U_duality}
\end{align}
We start with the $SU$-$U$ duality \eqref{eq:U_SU_duality}. We show that this duality is related to the $\mc{N}=2$ $SU$-$U$ duality discussed in \cite{Aharony:1997bx}. This duality was generalized to an arbitrary number of fundamentals and anti-fundamentals in \cite{Aharony:2014uya}. For the case at hand, we are interested in the duality where there are no antifundamentals (and $N_f<2k$):
\beq
SU(N)_k+N_f\; Q\longleftrightarrow U(N_f/2+k-N)_{-k,\frac12 N_f-N}+N_f\; q 
\eeq{eq:N=2_only_funds}
We argue that by deforming the $\mc{N}=2$ duality \eqref{eq:N=2_only_funds}, one can find the $\mc{N}=1$ duality \eqref{eq:U_SU_duality}. To see this, we first remember that the $\mc{N}=2$ fixed point is unstable to $\mc{N}=1$ $\Phi^4$ deformations at large $k$, so that such an $\mc{N}=1$ deformation should make the theory flow to a new $\mc{N}=1$ fixed point. 
Now, if we think of the $\mc{N}=2$ vector multiplet as an $\mc{N}=1$ vector multiplet with an additional $\mc{N}=1$ matter multiplet, then we can integrate out the matter multiplet at large $k$ (since its mass is proportional to $k$) and end up with an $\mc{N}=1$ duality. This integration out induces a $\grF^4$ coupling, which seems to have the correct sign to make us flow to the correct $\mc{N}=1$ fixed point shown in Figure \ref{fig:RGflow}. Furthermore, this integration out shifts the CS level, and we find exactly the $\mc{N}=1$ theories discussed above.

The picture we find is thus the following. It is possible that by just integrating out one-half of the $\mc{N}=2$ vector multiplet, we flow to the $\mc{N}=1$ duality, due to the appearance of the $\grF^4$ interaction. But even if this is not so, we can also add a $\grF^4$ term in addition to the integration out to make sure that we flow to the correct fixed point. We can thus flow from the $\mc{N}=2$ duality to the $\mc{N}=1$ duality.

Another piece of evidence for this relation is the phase diagrams for these theories. Consider the $\mc{N}=2$ theories discussed above. We can add real masses (and an FI parameter where relevant) to these theories, and find the phases as a function of these parameters close to the fixed point. Assuming we give an identical real mass to all of the matter fields, we find that the phases as a function of the real mass are identical to the phases we found for the $\mc{N}=1$ theories in Section \ref{phase_diagrams} around the fixed point. Since the RG flow is "short", we might expect the phases to be similar for these two theories. This again points to a relation between the $\mc{N}=2$ and the $\mc{N}=1$ dualities.

We can do an identical analysis for the first $U$-$U$ duality \eqref{eq:first_U_U_duality}. This time, the relevant $\mc{N}=2$ duality is Giveon-Kutasov duality \cite{Giveon:2008zn}. Once again, we must start by getting rid of the antifundamental matter fields in Giveon-Kutasov by adding mass terms. This was done in \cite{Benini:2011mf}, and the result is
\beq 
U(N)_k+N_f\;\grF\longleftrightarrow  U\left(N_f/2+k-N\right)_{-k}+N_f\;\widetilde{\grF}
\eeq{eq:first_UU_N2_duality}
We again find that a naive integration out of the $\mc{N}=1$ chiral multiplet embedded in the $\mc{N}=2$ vector multiplet immediately leads to the $\mc{N}=1$ duality described above. We thus expect a similar behavior in this case as well. We also find once again the the phases of the $\mc{N}=2$ theory as a function of the real mass match the phases of the $\mc{N}=1$ theory as a function of the mass $m$ (for the case where the mass preserves the global symmetries).

Finally, we discuss the second $U$-$U$ duality \eqref{eq:second_U_U_duality}. Interestingly, the relevant $\mathcal{N}=2$ duality has not yet appeared in the literature. This should be
\beq
U(N)_{k+N,k-N}+N_f\;\Phi,\tilde{\Phi}\longleftrightarrow U(N_f+k)_{-k-N,k-N+2N_f}+N_f\;\Psi,\tilde{\Psi}+N_f^2\;M 
\eeq{new_N=2}
where the $M$'s are gauge singlets and the RHS has the superpotential $W=\tilde{\Psi}M\Psi$. Following the procedure discussed above, this should flow to our $\mc{N}=1$ duality. We have performed some nontrivial checks on this $\mathcal{N}=2$ duality. First, one can match the phases of the two sides as a function of real and complex masses for the quarks. Second, using localization \cite{Kapustin:2009kz,Hama:2010av,Jafferis:2010un}, one can also match the partition functions of the two sides for the simpler cases of the duality.

We make two final comments. First, we note that the range of validity for the $\mc{N}=2$ theories directly implies the range of validity that were found for the $\mc{N}=1$ dualities above (for example, the range of the $\mc{N}=2$ $SU$-$U$ duality discussed in \cite{Benini:2011mf} directly leads to the range $k>-\min(N_f,N)$ for the $\mc{N}=1$ $SU$-$U$ duality). Since the range of the $\mc{N}=1$ dualities was found here independently, this is further evidence of the correspondence between the $\mc{N}=1$ and $\mc{N}=2$ dualities. Second, we note that the relation to $\mc{N}=2$ dualities can explain the fact that there is a single phase transition in the phase diagram to all orders in perturbation theory. Indeed, in the $\mc{N}=2$ case this fact is obvious; it results from the fact that the superpotential in these theories is holomorphic, and so the standard non-renormalization theorems apply. This means that the vacua are often given by the solutions of the classical (or 1-loop) F-term equations, which can easily be tuned to have such unnatural structures (and indeed this can be seen in the phase structure of the $\mc{N}=2$ theories described above when we add real masses to the matter fields). If the $\mc{N}=1$ and $\mc{N}=2$ theories are then close (in an RG flow sense), we might expect their phase diagrams to be similar as well. This might lead to there being a single phase transition in the $\mc{N}=1$ theory as well.

\section{Summary and Conclusions}
\setcounter{equation}{0}

In this work, we considered 2+1$d$ $\mc{N}=1$ supersymmetric gauge theories, with matter (mostly) in the fundamental representation. 
Using supergraphs, we have found the 1-loop effective superpotential for these theories with matter in an arbitrary representation. This superpotential allows us to find the phase diagram for these theories by solving the F-term equations.

Next, restricting ourselves to matter fields in the fundamental representation, we find a universal form for the solutions of the F-term equations and for the resulting phase diagrams for many different gauge groups (including $SO(N),O(N),$ $Sp(N),SU(N)$ and $U(N)$), see Figure \ref{fig:general_phase_diagram}. These phase diagrams have three regimes - two semiclassical phases for large $|m|$ with a single vacuum, and one intermediate phase with many SUSY vacua.
In particular, we find that there is a single point where the Witten index jumps (due to vacua appearing from infinity), and a single phase transition where many different vacua collide.

One can understand this universal form using a simple symmetry argument. This argument also proves that there is a single phase transition to all orders in perturbation theory. Thus, even though this result seems highly unnatural, it is actually an exact result. This fact (together with the phase diagrams we found) allows us to write down dualities relating the different theories, summarized in equations \eqref{intro_duality}-\eqref{intro_U_duality_2}. All of these dualities except one were shown to be related to well known $\mc{N}=2$ dualities (except for one, for which we conjecture the corresponding $\mc{N}=2$ duality \eqref{new_N=2}). 

These dualities lead to some interesting conclusions. In particular, we find many theories which have emergent global symmetries and supersymmetry in the IR. Furthermore, we find a series of theories which have emergent time-reversal symmetry in the IR, which might indicate that they have an exact quantum moduli space.

This paper includes both conjectures (for example, the dualities) and some exact results (for example, the single phase transition at $m=m_*$ where many vacua collide). Even more exact results have appeared recently in the literature for theories with $\mc{N}=1$ SUSY. The fact that one can find exact results in theories with $\mc{N}=1$ might seem very surprising at first; As mentioned in the introduction, the superpotentials in these theories are not holomorphic, and so they do not possess the full power of supersymmetry. However, our results seem to indicate that some of the exact results that appear in this paper are still a direct result of the holomorphy of the superpotential in $\mc{N}=2$ SUSY. In other words, most of the exact results can be understood by deforming theories with $\mc{N}=2$ SUSY. It would be interesting to see if other exact results in $\mc{N}=1$ theories can also be understood directly from $\mc{N}=2$ theories\footnote{In particular, it would be interesting to understand whether the exact moduli spaces appearing in \cite{Gaiotto:2018yjh} can be understood using deformations of $\mc{N}=2$ theories, or whether they are intrinsically $\mc{N}=1$ phenomena. On one hand, it is very common to have exact moduli spaces in theories with a holomorphic superpotential, and so this doesn't sound unreasonable. On the other hand, this seems highly nontrivial, since (for example) moduli spaces of theories with holomorphic superpotentials are K\"{a}hler manifolds, while the moduli spaces that were found for $\mc{N}=1$ theories were not K\"{a}hler.}. Furthermore, it would be interesting to see whether we can continue this flow to find exact results in non-SUSY theories as well. As we have mentioned, the connection between the non-SUSY bosonization dualities and some $\mc{N}=2$ dualities has already been studied in the literature (and a similar connection to our $\mc{N}=1$ dualities should also exist). Hopefully, the recent results in theories with $\mc{N}=1$ SUSY might lead to new results in non-SUSY theories as well.

\section*{Acknowledgements}
We are happy to thank Zohar Komargodski for numerous discussions and guidance. MR is happy to thank Warren Siegel for useful discussions and Ulf Lindstr\"om for his help at the beginning of this project. CC is happy to thank Yoonjoo Kim for explaining some relevant mathematics. MR was supported in part by NSF Grant \#~PHY1620628.
AS is grateful for the hospitality of the Simons Center for Geometry and Physics, Stony Brook University, where this collaboration began. AS is supported by an Israel Science Foundation center for excellence grant and by the I-CORE program
of the Planning and Budgeting Committee and the Israel Science Foundation (grant
number 1937/12). 

\begin{appendices}
\addtocontents{toc}{\protect\setcounter{tocdepth}{1}}

\section{Superspace Conventions}\label{app_superspace_conventions}
\setcounter{equation}{0}
We follow \cite{Gates:1983nr}. Spinor indicies are raised and lowered with 
\beq
C_\ab=-C^\ab=  (\sigma_2)_\ab  \then C_\ab C^{\ga\de}=\de_\al{}^\de\de_\be{}^\de-\de_\be{}^\de\de_\al{}^\de~~.
\eeq{Cab}
We use the conventions that for a spinor 
\beq
\psi^\al=C^\ab\psi_\be~~,~~\psi_\be =\psi^\al C_\ab~~,~~\psi^2=\frac12\psi^\al\psi_\al=i\psi^+\psi^-\
\eeq{psic}
Spinor derivatives obey the usual algebra:
\beq
\{D_\al,D_\be\}=2i\pa_\ab~~.
\eeq{Dalg}
The convention is again $D^2=\frac12D^\al D_\al$. These identities follow trivially:
\beq
D_\al D_\be =i\pa_\ab -C_\ab D^2~~,~~ D^\al D_\be D_\al =0~~,~~ D^2 D_\al =- D_\al D^2 = i \pa_\ab D^\be~~,
\eeq{Did}
as well as
\beq
\pa^\ab\pa_{\ga\be}=\de_\ga{}^\al \square ~~,~~(D^2)(D^2)=\square~~,~~ \square \equiv \frac12 \pa^\ab\pa_\ab~~.
\eeq{Bid}

The free scalar superfield action is
\beq
S_X=\frac12 \int d^2\ta ~(XD^2 X +m X^2)=\frac12 \int d^2\ta \left(-\frac12D^\al X D_\al X +mX^2\right)~~.
\eeq{SX}
The gauge multiplet is described by covariantizing the spinor and vector derivatives:
\beq
\{\na_\al,\na_\be\}=2i\na_\ab~~,~~ \na_\al\equiv D_\al -i \G_\al~~,
\eeq{YM}
where the generators are hermitian (which is why there is an $i$ in the definition of the gauge covariant derivative $\na_\al$).
The Bianchi identities imply:
\beq
[\na_\ga,\na_\ab ] = C_{\ga\al}W_\be + C_{\ga\be}W_\al ~~,~~~ \na^\al W_\al  = 0~~.
\eeq{Bianchi}
The explicit form of $W_\al$ is
\beq
W_\al = \frac{i}6[\na^\be,\{\na_\be,\na_\al\}]=\frac12 \left( D^\be D_\al \G_\be -i[\G^\be,D_\be\G_\al]-\frac13[\G^\be,\{\G_\be,\G_\al\}] \right)~~.
\eeq{Curv}

\section{Separating the Massive YM mode from the CS mode}\label{appendix YM and CS}
Consider the Higgs mechanism in a YM-CS theory. The gauge field then has two modes \cite{Dunne:1998qy}; at large CS level $k$, we find that one is very massive (with mass of order $m\sim kg^2$) while the mass of the other field is very small. Here we show how to separate these modes to integrate out only the very massive mode.

We can replace $S_{WW}$ with a simpler action by introducing an extra spinor superfield $\psi^\al$:
\beq
S_{\psi\psi}=\int d^2\ta ~  \hbox{Tr}\left(-\frac{g^2k^2}{2\pi^2}\psi^2
-\frac{k}{\pi}\psi^\al W_\al\right)=-\frac{g^2k^2}{2\pi^2}\int d^2\ta ~\hbox{Tr}\left(\frac12\psi^\al\psi_\al
+\frac{2\pi}{g^2k}\psi^\al W_\al\right)
\eeq{Spsi}
Note that eliminating $\psi$ gives back $S_{WW}$.
We now cancel the $\psi^\al W_\al$ term by shifting $\G$:
\beq
\G=\G'+\psi ~~\Rightarrow~~\na=\na'-i\psi~, 
\eeq{gshift}
Using (\ref{Curv}), we have:
\beq
W_\al = W'_{\al}+\frac12 \left( \na'{}^\be \nabp\al \psi_\be -i[\psi^\be,\nabp\be\psi_\al]-\frac13[\psi^\be,\{\psi_\be,\psi_\al\}] \right)~~.
\eeq{psiCurv}
We find the shifted Chern-Simons term directly from (\ref{CSs}):
\beq
S_{CS}[\G'\!+\!\psi]=S_{CS}[\G']+\frac{k}{\pi}\int d^2\ta ~\hbox{Tr}\left(\!\psi^\al W'_{\al}+\frac14\left[\psi^\al\na'{}^\be \nabp\al \psi_\be -\frac{2i}3\{\psi^\al,\psi^\be\}\na'_\al\psi_\be-\frac16\{\psi^\al,\psi_\be\}\{\psi_\al,\psi_\be\}\right]\right)
\eeq{CSpsi}
Notice that the $\psi^\al W'_\al$ cross-term cancels when we add $S_{\psi\psi}+S_{CS}[\G'\!+\!\psi]$; using (\ref{psiCurv}), we find\cite{karlhede1987supersymmetric}
\beq 
S_{\G'\psi}=S_{CS}[\G']-\frac{k}{2\pi} \int d^2\ta~\hbox{Tr}\left(\frac12\psi^\al\na'{}^\be \nabp\al \psi_\be-\frac{2i}3\{\psi^\al,\psi^\be\}\na'_\al\psi_\be-\frac14\{\psi^\al,\psi_\be\}\{\psi_\al,\psi_\be\}+\frac{g^2k}\pi\psi^2\right)
\eeq{spsigp}
The superfield $\psi$ thus has a mass $\kappa=\frac{kg^2}{2\pi}$ (before corrections due to the Higgs mechanism), and we identify it with the very massive YM mode. We can now integrate out this mode to find the effective action for the remaining fields. 

After performing this separation, since the gauge field $\G'$
has only a Chern-Simons kinetic term, gauge-fixing is simpler, and no Nielsen-Kallosh ghosts such as those described in section \ref{gauge_theories_in_superspace} are needed; the gauge fixing term can be chosen to be, e.g.,
\beq
S_{fix}^0= -\frac{k}{4\pi\al}\int d^2\ta~\hbox{Tr}(D^\al\G_\al)^2
\eeq{fixC}
or the usual $R-\xi$ gauge, etc.

\section{Superpotential for \texorpdfstring{$U(N)_{k,k'}$}{unkk}} \label{app_superpotential_UN}
\setcounter{equation}{0}
We find the form of the superpotential for a $U(N)_{k,k'}$ gauge theory. Let us redo the calculation from Section \ref{massless_1loop_superpotential}. This time, the calculation is more complicated, since we have two different propagators (for $k$ and $k'$).

Let us discuss the form of the propagator. Note that we can rewrite the propagator corresponding to level $k$ (ignoring group factors)
\begin{equation}
\begin{aligned}
\Delta_\al^{\;\;\be}(\kappa)&=-g^2\frac{\delta_\al^{\;\;\be}\left(\kappa D^2+p^2\right)+\left(\kappa-D^2\right)p_\al^{\;\;\be}}{p^2\left(\kappa^2+p^2\right)}
\\
&=-g^2\frac{\delta_\al^{\;\;\be}\left(\kappa D^4+p^2D^2\right)D^{-2}+\left(\kappa-D^2\right)p_\al^{\;\;\be}}{p^2\left(\kappa^2+p^2\right)}
\\
&=-g^2\frac{\kappa-D^2}{\kappa^2+p^2}\frac{-\delta_\al^{\;\;\be}p^2D^{-2}+p_\al^{\;\;\be}}{p^2}
\end{aligned}
\end{equation}
We can now relate the propagators with different $k$'s. Again, we use the trick \ref{eq:imaginary_trick}, which amounts to replacing $D^2$ by $ip$ and extracting the imaginary part at the end. Thus $U(1)$ and $SU(N)$ propagators are then related by
\begin{align*}
\Delta_\al^{\;\;\be}(\tilde{\kappa})&=
\frac{\tilde{\kappa}-ip}{\tilde{\kappa^2}+p^2}
\frac{\kappa^2+p^2}{\kappa-ip}
\Delta_\al^{\;\;\be}(\kappa)
\\
&=\frac{\kappa+ip}{\tilde{\kappa}+ip}\Delta_\al^{\;\;\be}(\kappa)
\\
&\equiv R\Delta_\al^{\;\;\be}(\kappa)
\end{align*}

We must now re-sum the diagrams. We reintroduce the $\grf^2=\bar{\grf}T^{(a}T^{b)}\grf$ factor, which multiplies each propagator. We observe that each $U(1)$ propagator comes with two factors of $T^0$. We can thus absorb the factor of $R$ into the generator $T^0$ using $\tilde{T}^0=\sqrt{R}T^0$, The propagators are now all equal, but the new definition for $\grf^2$ is
\beq
\tilde{\phi}^2\equiv \phi^\dagger \tilde{T}^{(a} \tilde{T}^{b)} \phi, ~~~\tilde{T}^0=RT^0, ~\tilde{T}^a=T^a,~ a \in SU(N)
\eeq{phiid}
The following part of the calculation proceeds as before. We can re-sum the diagrams to find the superpotential:
\begin{align*}
W&=-\int\frac{d^3p}{(2\pi)^3}\frac{1}{|p|}\mbox{Im}\Tr \log \left(1+\frac{g^2\tilde{\phi}^2(i\kappa |p|+p^2)}{p^2(\kappa^2+p^2)}\right)
\\
&=-\int\frac{d^3p}{(2\pi)^3}\frac{1}{|p|}\mbox{Im}\Tr \log \left(1+\frac{g^2\tilde{\phi}^2}{(-i\kappa |p|+p^2)}\right)
\end{align*}
Note that this integral is still much more complicated than before, since $\tilde{\grf}$ depends on $p$. However, this form allows us to perform numerical calculations to find the vacua for the theory.

\section{Matching the Witten Index}\label{app_matching_witten_index}
\setcounter{equation}{0}
\subsection{Generalities}\label{witten index generalities}
In this appendix we study the Witten index of the phase diagrams for $SU(N)$ and $U(N)$ gauge theories described in Sections \ref{phases_of_SUN},\ref{phases_of_UN}. We show that for these theories, the Witten index jumps only once (at $m=0$), and does not jump at the phase transition point $m=m_*$. This is consistent with the fact that vacua appearing from infinity can cause the Witten index to jump. We also find that as required, the Witten index of the vacua coming in from infinity exactly matches the expected jump

We focus on the calculation of the Witten index for the large positive mass phase and for the intermediate phase in Figure \ref{fig:general_phase_diagram}. We show that the Witten index of both phases is equal, so that there is no jump at $m=m_*$. This immediately leads to the necessity of a jump at $m=0$, since a simple calculation shows that the Witten indices for positive mass and negative mass are not equal in general.

The following facts about the Witten index are used. First, the Witten index for a SUSY NLSM is equal to the Euler characteristic of the target manifold $\chi(\mathcal{M})$ \cite{Witten:1982df}.
In particular, for the Grassmannian Manifold $\mathcal{M}_{n,m}=\frac{SU(N_f)}{S[U(N_f-m)\times U(m)]}$, we find\footnote{One way to obtain this result is using Atiyah-Bott Localization Theorem \cite{hori2003mirror}, since the fixed points under the action of $U(1)^n$ on $\mathcal{M}_{n,k}$ are just the $\begin{pmatrix}n\\k\end{pmatrix}$ 
diagonal elements.}
\begin{equation}
\label{eq:euler}
I(\mathcal{M}_{n,m})=\chi(\mathcal{M}_{n,m})=\begin{pmatrix}n\\m\end{pmatrix}
\end{equation}
We also make extensive use of the Witten indices of the following $\mc{N}=1$ vector multiplets:
\begin{equation}\label{eq:wittenSU}
I(SU(N)_{k}^{\mc{N}=1})=\begin{cases}
\binom{\frac{N}{2}+k-1}{N-1} & k>0\\
0 & -N/2<k\leq 0
\end{cases}
\end{equation}
\begin{equation}\label{eq:wittenU}
I(U(N)_{k+N/2,k}^{\mc{N}=1})=\begin{cases}
\binom{N+k-1}{k-1} & k > 0\\
0 & -N/2<k\leq 0
\end{cases}
\end{equation}
It is believed that these theories break SUSY dynamically when their Witten index vanishes. 

It is important to note that $I(U(N)_{N/2,0}^{\mc{N}=1})=0$. This is because the $U(1)$ part of the theory has a circle of vacua, parametrized by the dual photon. Since the $\chi(S^1)=0$, we find that the Witten index of this theory vanishes. This fact was also used in \cite{Bashmakov:2018wts}.

Vandermonde's Identity also plays important role:
\begin{equation}\label{eq:vandermonde}
\begin{aligned}
\sum_{n=0}^{K'} \begin{pmatrix}M\\n\end{pmatrix}\begin{pmatrix}K\\K'-n\end{pmatrix}
=\begin{pmatrix}K+M\\K'\end{pmatrix}
\end{aligned}
\end{equation}

Finally, a comment about the sign ambiguity of the Witten index is in order. There is a sign ambiguity for the operator $(-1)^F$ in finite volume \cite{Witten:1982df, Witten:1999ds}. The overall sign is purely determined by our own choice for the Hilbert space, and we set it as positive for large positive $k$ for both $SU(N)_k$ and $U(N)_k$.

\subsection{\texorpdfstring{$SU(N)_{k+N/2}$}{sunkn} with \texorpdfstring{$N>N_f$}{N>nf}}
\label{app_SUN_N>Nf}
We study the theories described in Section \ref{SUN_Nf_smaller_N}. We show that the Witten index does not jump across the phase transition at $m=m_*$. As a result, we find that it must jump at $m=0$, when the new vacua appear from infinity.

\subsubsection{\texorpdfstring{$k\geq N_f/2$}{k>}}
There are no SUSY-breaking vacua in this case. For small positive mass, the sum of the Witten indices for all of the vacua is (using \eqref{eq:wittenSU},\eqref{eq:vandermonde}):
\beq \sum_{n=0}^{N_f}I(\mc{M}_{N_f,n}\times SU(N-n)_{k+N/2-N_f/2+n/2}^{\mc{N}=1})=\sum_{n=0}^{N_f} \begin{pmatrix}N_f\\n\end{pmatrix}\begin{pmatrix}
N+k-\frac{N_f}{2}-1\\N-n-1\end{pmatrix}=
\begin{pmatrix}N+k+\frac{N_f}{2}-1\\N-1\end{pmatrix} 
\eeq{c2id}
which is precisely the Witten index of the large positive mass phase, which is an $\mc{N}=1$ $ SU(N)_{k+N/2+N_f/2}$ vector multiplet (see \eqref{eq:wittenSU}).

\subsubsection{\texorpdfstring{$-N_f/2 \leq k < N_f/2$}{<k<}}
Here, some vacua in the range $0<m<m_*$ become SUSY-breaking (as discussed in Section \ref{free_N1_vector_multiplets}), and so they do not contribute to the index. The total Witten index in this range becomes
\begin{align}
\sum_{n=0}^{N_f}I(\mc{M}_{N_f,n}\times SU(N-n)_{k+N/2-N_f/2+n/2}^{\mc{N}=1}) =&
\text{$
\sum_{n=0}^{N_f/2-k-1}I(\cancelto{ \text{\footnotesize{SUSY-breaking}}}{\mc{M}_{N_f,n}\times SU(N-n)_{k+N/2-N_f/2+n/2}^{\mc{N}=1}})$}\nonumber
\\  &+\sum_{n=N_f/2-k}^{N_f}I(\mc{M}_{N_f,n}\times SU(N-n)_{k+N/2-N_f/2+n/2}^{\mc{N}=1})\nonumber\\
=&\sum_{n=0}^{N_f/2+k} \begin{pmatrix}N_f\\n\end{pmatrix}\begin{pmatrix}
N+k-\frac{N_f}{2}-1\\k+\frac{N_f}{2}-n\end{pmatrix}\nonumber\\
=&\begin{pmatrix}N+k+\frac{N_f}{2}-1\\N-1\end{pmatrix} 
\end{align}
Which is precisely the Witten index of the large positive mass phase, an $\mc{N}=1$ $ SU(N)_{k+N/2+N_f/2}$ vector multiplet.

\subsubsection{\texorpdfstring{$-N/2 \leq k < -N_f/2$}{-n<k}}
All of the vacua break SUSY in this region since $\vert k+N/2 \pm N_f/2 \vert ~<~ N/2$. Thus the Witten index vanishes on both sides of the transition.

\subsection{\texorpdfstring{$SU(N)_{k+N/2}$}{sunkpn} with \texorpdfstring{$N\leq N_f$}{nleq}}
\label{app_SUN_N<Nf}

\subsubsection{\texorpdfstring{$k\geq N_f/2$}{kgeq}}\label{SUN_N_larger_Nf_k_larger_Nf}

There are no SUSY-breaking vacua in this case. For small positive mass, the sum of the Witten indices for all of the vacua is:
\begin{align} 
I&=\cancelto{0}{I(\frac{U(N_f)}{SU(N)\times U(N_f-N)})}+\sum_{n=0}^{N-1} I(\mathcal{M}_{N_f,n} \times SU(N-n)_{k+N/2-(N_f-n)/2}^{\mc{N}=1})\nonumber\\
&=\sum_{n=0}^{N-1} \begin{pmatrix}N_f\\n\end{pmatrix}\begin{pmatrix}N-\frac{N_f}{2}+k-1\\N-1-n\end{pmatrix}\nonumber\\
&=\begin{pmatrix}N+\frac{N_f}{2}+k-1\\N-1\end{pmatrix}
\end{align}
which is the Witten index of the large positive mass phase, an $\mc{N}=1$ $ SU(N)_{k+N/2+N_f/2}$ vector multiplet. The first term vanishes because it is the Euler characteristic of an odd-dimensional manifold.

\subsubsection{\texorpdfstring{$-N+N_f/2 < k < N_f/2$}{-npnf}}
SUSY-breaking occurs for the phases with $0\leq n <N_f/2-k$, and so they do not contribute to the index. The full index for $0<m<m_*$ is:
\begin{align}
\sum_{n=0}^{N-1}I(\mc{M}_{N_f,n}\times SU(N-n)_{k+N/2-(N_f-n)/2}^{\mc{N}=1})=&
\text{$\cancelto{0}{I(\frac{U(N_f)}{SU(N)\times U(N_f-N)})}$}\nonumber\\
&+
\text{$\sum_{n=0}^{N_f/2-k-1}I(\cancelto{ \text{\footnotesize{SUSY-breaking}}}{\mc{M}_{N_f,n}\times SU(N-n)_{k+N/2-(N_f-n)/2}^{\mc{N}=1}})$}\nonumber\\
&+\text{$\sum_{n=N_f/2-k}^{N-1}I(\mc{M}_{N_f,n}\times SU(N-n)_{k+N/2-(N_f-n)/2}^{\mc{N}=1})$}\nonumber\\
=&
\sum_{n=0}^{N+k-\frac{N_f}{2}-1} \begin{pmatrix}N_f\\ \frac{N_f}{2}+k-n\end{pmatrix}\begin{pmatrix}
N+k-\frac{N_f}{2}-1\\n\end{pmatrix}\nonumber\\
=&\begin{pmatrix}N+k+\frac{N_f}{2}-1\\N-1\end{pmatrix} 
\end{align}
and again we obtain the Witten index of the large positive mass phase, an $\mc{N}=1$ $ SU(N)_{k+N/2+N_f/2}$ vector multiplet. As above, the first term vanishes because it is the Euler characteristic of an odd-dimensional manifold.

\subsubsection{\texorpdfstring{$-N/2 \leq k\leq -N+N_f/2$}{-nleq}}
In this case, there are no SUSY breaking for all phases, since $k+N/2-(N_f-n)/2 \leq -(N-n)/2$ and $k+N/2+N_f/2\geq N/2$. The calculation is similar to \ref{SUN_N_larger_Nf_k_larger_Nf}.

\subsection{\texorpdfstring{$U(N)_{k+N/2,k}$ with $ N>N_f$}{unknp}}
\label{app_UN_N>Nf}

\subsubsection{\texorpdfstring{$k>N_f/2$}{kgnf}}
The total index for small positive $m$ is (using  \eqref{eq:wittenU}):
\begin{align}
 &\sum_{n=0}^{N_f}I(      \mathcal{M}_{N_f,n} \times U(N-n)_{k+N/2-N_f/2+n/2,k-N_f/2+n}^{\mc{N}=1})
=\sum_{n=0}^{N_f} 
\begin{pmatrix}N_f\\N_f-n\end{pmatrix}\begin{pmatrix}
N+k-\frac{N_f}{2}-1\\k-\frac{N_f}{2}+n-1\end{pmatrix}
=\begin{pmatrix}
N+k+\frac{N_f}{2}-1\\N
\end{pmatrix}
\end{align}
which is the index of the large positive mass phase, an $\mc{N}=1$ $U(N)_{k+N/2+N_f/2,k+N_f/2}$ vector multiplet.

\subsubsection{\texorpdfstring{$-N_f/2 \leq k\leq N_f/2$}{-nf2leq}}\label{app_why_index_vanishes_for_Uk0}

Here, some vacua are SUSY-breaking, and so they do not contribute to the index. The total Witten index for small positive mass is:
\begin{align}\label{d10}
\sum_{n=0}^{N_f}I(\mathcal{M}_{N_f,n} \times U(N-n)_{k+N/2-N_f/2+n/2,k-N_f/2+n}^{\mc{N}=1})
=&
\text{$
\sum_{n=0}^{N_f/2-k-1}I(\cancelto{\text{\footnotesize{SUSY-breaking}}}{U(N-n)_{k+N/2-N_f/2+n/2,k-N_f/2+n}^{\mc{N}=1}})$}\nonumber\\
&+
I(\text{$\cancelto{0}{U(N-N_f/2+k)_{k/2+N/2-N_f/4,0}^{\mc{N}=1}} $}) \nonumber\\
&+
\text{$
\sum_{n=N_f/2-k+1}^{N_f}I(\mc{M}_{N_f,n}\times U(N-n)_{k+N/2-N_f/2+n/2,k-N_f/2+n}^{\mc{N}=1})$}
\nonumber\\
=& \sum_{n=N_f/2-k+1}^{N_f} \begin{pmatrix}N_f\\N_f-n\end{pmatrix}\begin{pmatrix}
N+k-\frac{N_f}{2}-1\\k-N_f/2+n-1\end{pmatrix}\nonumber\\
=&\begin{pmatrix}N+k+\frac{N_f}{2}-1\\N\end{pmatrix} 
\end{align}
which again, matches with the index of the large positive m vacuum. As explained below \eqref{eq:wittenU}, the Witten index of the term on the second line vanishes: generally, an $\mathcal{N}=1$ $U(N)_{N/2,0}$ vector multiplet leads to a $U(N)_0$ TQFT in the IR. The $U(1)$ part of this TQFT has a circle of vacua, parametrized by the dual scalar. The Euler characteristic of a circle is zero, and so the Witten index of this theory vanishes.

\subsubsection{\texorpdfstring{$-N/2\leq k \leq -N_f/2 $}{-n2leqkleq}}

In this region, all phases break supersymmetry because $| k+N/2 \pm N_f/2 | < N/2$. Thus the Witten index matches trivially.

\subsection{\texorpdfstring{$U(N)_{k+N/2,k}$}{unnn} with \texorpdfstring{$N\leq N_f$}{nnfleq}}\label{app_UN_N<Nf}

\subsubsection{\texorpdfstring{$k\geq N_f/2$}{kgeqnf}}
All of the small positive mass vacua are SUSY-preserving, and so the total index for small positive $m$ is:
\begin{align*}
 &\sum_{n=0}^{N}I(      \mathcal{M}_{N_f,n} \times U(N-n)_{k+N/2-N_f/2+n/2,k-N_f/2+n}^{\mc{N}=1})
=
\sum_{n=0}^{N} 
\begin{pmatrix}N_f\\n\end{pmatrix}\begin{pmatrix}
N+k-\frac{N_f}{2}-1\\N-n\end{pmatrix}
=
\begin{pmatrix}
N+k+\frac{N_f}{2}-1\\N
\end{pmatrix}
\end{align*}
which is again consistent with the index for large positive $m$.

\subsubsection{\texorpdfstring{$-N+N_f/2 <k<N_f/2$}{-nnf2knf2}}

Here, some of the vacua can break SUSY. The total Witten index for the small positive mass phase is:
\begin{align}
\sum_{n=0}^{N}I(      \mathcal{M}_{N_f,n} \times U(N-n)_{k+N/2-N_f/2+n/2,k-N_f/2+n}^{\mc{N}=1})
=&
\text{$
\sum_{n=0}^{N_f/2-k-1}I(\cancelto{\text{\footnotesize{SUSY-breaking}}}{U(N-n)_{k+N/2-N_f/2+n/2,k-N_f/2+n}^{\mc{N}=1}})$}\nonumber\\
&+
I(\text{$\cancelto{0}{U(N-N_f/2+k)_{k/2+N/2-N_f/4,0}^{\mc{N}=1}} $})\nonumber\\
&+
\text{$
\sum_{n=N_f/2-k+1}^{N}I(\mc{M}_{N_f,n}\times U(N-n)_{k+N/2-N_f/2+n/2,k-N_f/2+n}^{\mc{N}=1})$}\nonumber\\
&=
\sum_{n=N_f/2-k+1}^{N} \begin{pmatrix}N_f\\n\end{pmatrix}\begin{pmatrix}
N+k-\frac{N_f}{2}-1\\N-n\end{pmatrix}\nonumber\\
&=
\begin{pmatrix}N+k+\frac{N_f}{2}-1\\N\end{pmatrix} 
\end{align}
which is again identical to the index of the vacuum at large positive $m$. The Witten index of the second line vanishes as explained below \eqref{eq:wittenU},\eqref{d10}.

\section{Proof That There Is Only One Phase Transition}\label{app_one_phase_transition}
\setcounter{equation}{0}
We prove that the solutions $\grf_n=\begin{pmatrix}
v_n(m)\cdot \id_{n\times n} & \\
& 0
\end{pmatrix}$ must all coalesce at the same value of $v$ and $m$, which proves that there is only one phase transition. Our arguments are similar to those used in Landau-Ginzburg theory. The main logic is the following: consider some effective potential $U$ for small $\grf$. This can be expanded in $\grf$, and we find $U(\grf)=\grl_1\grf^2+\grl_2\grf^4+...$ for some constants $\grl_1,\grl_2$ (we assume $\grl_2>0$). We find that this model has a phase transition point, which occurs when $\grl_1=0$. We can now repeat this procedure for the effective potential $W^{(n)}$ of a specific solution $\grf_n$. What we show is that for any $\grf_n$ we have $\grl_1^{(n)}=n\grl$, where $\grl$ is independent of $n$. The phase transition for all $\grf_n$'s is thus at the same point $\grl=0$.

We start by noticing that a phase transition (i.e. two of the solutions $\phi_n,\phi_k$ colliding) can only occur at $\phi=0$. This is obvious, since to have $\phi_n=\phi_k$ we must have $v_n=v_k=0$. It is thus enough to show that all of the phase transitions occur at the same mass $m_*$. Consider one solution $\grf_n$ and assume that it undergoes a phase transition at $m_*^{(n)}$ (in other words, $v_n\left(m_*^{(n)}\right)=0$) Assume that the phase transition that occurs for the smallest $m$ includes the solution $v_n$, and denote this $m$ by $m=m_n^*$. Since $v_n$ must vanish at the phase transition, we find $v_n(m=m_n^*)=0$.
  
Now we expand the superpotential around the phase transition point $\grf=0$:
\beq 
W(m)=W_0(m)+W_2(m)\tr \phi_n^2+(W_4(m)\tr \phi_n^4+\tilde{W}_4(m)(\tr \phi_n^2)^2)+... 
\eeq{wexpand}
Plug in the solution $\grf_n$:
\beq 
W(m)=W_0(m)+W_2(m)nv_n^2+(W_4(m)nv_n^4+\tilde{W}_4(m)(nv_n^2)^2)+... 
\eeq{wexpand2}
This must solve the equation $\frac{\pa}{\pa v_n} W=0$, and so we have
\beq
0=2W_2(m)nv_n+4nv_n^3(W_4(m)+\tilde{W}_4(m)n)+... 
\eeq{wprime0}
Let us remove the solution at $v_n= 0$:
\beq 
0=2W_2(m)n+4nv_n^2(W_4(m)+\tilde{W}_4(m)n)+... 
\eeq{wprim01}
Finally, we plug in $m=m_n^*$ and  use the fact that $v_n(m_*)=0$ to find:
\beq 
0=W_2(m_n^*)n 
\eeq{w0primsol}
We thus find that the phase transition point $m_n^*$ is defined by the equation $W_2(m_n^*)=0$. Note that this equation does not depend on $n$, and so repeating this argument for any other $v_i$ gives the same $m_*$. We thus find that all other solutions $v_i$ must also vanish at this $m_*$, so that all of the solutions coalesce simultaneously. We conclude that there is only one phase transition.

\end{appendices}

\bibliographystyle{utphys}

\begin{thebibliography}{32}


\bibitem{Bashmakov:2018wts}
V.~Bashmakov, J.~Gomis, Z.~Komargodski, and A.~Sharon, ``{Phases of $\mathcal{N}{=}1$ Theories in 2+1 Dimensions},''
\href{http://arxiv.org/abs/1802.10130}{{\ttfamily arXiv:1802.10130 [hep-th]}}.

\bibitem{Benini:2018umh}
F.~Benini and S.~Benvenuti, ``{$\mathcal{N}{=}1$ dualities in 2+1
	dimensions},''
\href{http://arxiv.org/abs/1803.01784}{{\ttfamily arXiv:1803.01784 [hep-th]}}.

\bibitem{Gomis:2017ixy}
J.~Gomis, Z.~Komargodski, and N.~Seiberg, ``{Phases Of Adjoint QCD$_3$ And
	Dualities},''
\href{http://arxiv.org/abs/1710.03258}{{\ttfamily arXiv:1710.03258 [hep-th]}}.

\bibitem{Gaiotto:2018yjh}
D.~Gaiotto, Z.~Komargodski, and J.~Wu, ``{Curious Aspects of Three-Dimensional
	${\cal N}=1$ SCFTs},''
\href{http://arxiv.org/abs/1804.02018}{{\ttfamily arXiv:1804.02018 [hep-th]}}.

\bibitem{Eckhard:2018raj}
J.~Eckhard, S.~Schäfer-Nameki, and J.-M. Wong, ``{An $\mathcal{N}=1$ 3d-3d
	Correspondence},''
\href{http://arxiv.org/abs/1804.02368}{{\ttfamily arXiv:1804.02368 [hep-th]}}.

\bibitem{Inbasekar:2015tsa}
K.~Inbasekar, S.~Jain, S.~Mazumdar, S.~Minwalla, V.~Umesh, and S.~Yokoyama,
``{Unitarity, crossing symmetry and duality in the scattering of $
	\mathcal{N}=1 $ susy matter Chern-Simons theories},''
\href{http://dx.doi.org/10.1007/JHEP10(2015)176}{{\em JHEP} {\bfseries 10}
	(2015) 176},
\href{http://arxiv.org/abs/1505.06571}{{\ttfamily arXiv:1505.06571 [hep-th]}}.

\bibitem{Benini:2018bhk}
F.~Benini and S.~Benvenuti, ``{$N=1$ QED in 2+1 dimensions: Dualities and
	enhanced symmetries},''
\href{http://arxiv.org/abs/1804.05707}{{\ttfamily arXiv:1804.05707 [hep-th]}}.

\bibitem{Witten:1982df}
E.~Witten, ``{Constraints on Supersymmetry Breaking},''
\href{http://dx.doi.org/10.1016/0550-3213(82)90071-2}{{\em Nucl. Phys.}
	{\bfseries B202} (1982) 253}.

\bibitem{PhysRevD.7.1888}
S.~Coleman and E.~Weinberg, ``Radiative corrections as the origin of
spontaneous symmetry breaking,''
\href{http://dx.doi.org/10.1103/PhysRevD.7.1888}{{\em Phys. Rev. D}
	{\bfseries 7} (Mar, 1973) 1888--1910}.
\url{https://link.aps.org/doi/10.1103/PhysRevD.7.1888}.

\bibitem{Komargodski:2017dmc}
Z.~Komargodski, A.~Sharon, R.~Thorngren, and X.~Zhou, ``{Comments on Abelian
	Higgs Models and Persistent Order},''
\href{http://arxiv.org/abs/1705.04786}{{\ttfamily arXiv:1705.04786 [hep-th]}}.

\bibitem{AVDEEV1992561}
L.~Avdeev, G.~Grigoryev, and D.~Kazakov, ``Renormalizations in abelian
chern-simons field theories with matter,''
\href{http://dx.doi.org/https://doi.org/10.1016/0550-3213(92)90659-Y}{{\em
		Nuclear Physics B} {\bfseries 382} no.~3, (1992) 561 -- 580}.
\url{http://www.sciencedirect.com/science/article/pii/055032139290659Y}.

\bibitem{avdeev1993renormalizations}
L.~Avdeev, D.~Kazakov, and I.~Kondrashuk, ``Renormalizations in supersymmetric
and nonsupersymmetric non-abelian chern-simons field theories with matter,''
{\em Nuclear Physics B} {\bfseries 391} no.~1-2, (1993) 333--357.

\bibitem{Jain:2013gza}
S.~Jain, S.~Minwalla, and S.~Yokoyama, ``{Chern Simons duality with a
	fundamental boson and fermion},''
\href{http://dx.doi.org/10.1007/JHEP11(2013)037}{{\em JHEP} {\bfseries 11}
	(2013) 037},
\href{http://arxiv.org/abs/1305.7235}{{\ttfamily arXiv:1305.7235 [hep-th]}}.

\bibitem{Benini:2017aed}
F.~Benini, ``{Three-dimensional dualities with bosons and fermions},''
\href{http://dx.doi.org/10.1007/JHEP02(2018)068}{{\em JHEP} {\bfseries 02}
	(2018) 068},
\href{http://arxiv.org/abs/1712.00020}{{\ttfamily arXiv:1712.00020 [hep-th]}}.

\bibitem{Giveon:2008zn}
A.~Giveon and D.~Kutasov, ``{Seiberg Duality in Chern-Simons Theory},''
\href{http://dx.doi.org/10.1016/j.nuclphysb.2008.09.045}{{\em Nucl. Phys.}
	{\bfseries B812} (2009) 1--11},
\href{http://arxiv.org/abs/0808.0360}{{\ttfamily arXiv:0808.0360 [hep-th]}}.

\bibitem{Aharony:1997bx}
O.~Aharony, A.~Hanany, K.~A. Intriligator, N.~Seiberg, and M.~J. Strassler,
``{Aspects of N=2 supersymmetric gauge theories in three-dimensions},''
\href{http://dx.doi.org/10.1016/S0550-3213(97)00323-4}{{\em Nucl. Phys.}
	{\bfseries B499} (1997) 67--99},
\href{http://arxiv.org/abs/hep-th/9703110}{{\ttfamily arXiv:hep-th/9703110
		[hep-th]}}.

\bibitem{Benini:2011mf}
F.~Benini, C.~Closset, and S.~Cremonesi, ``{Comments on 3d Seiberg-like
	dualities},'' \href{http://dx.doi.org/10.1007/JHEP10(2011)075}{{\em JHEP}
	{\bfseries 10} (2011) 075},
\href{http://arxiv.org/abs/1108.5373}{{\ttfamily arXiv:1108.5373 [hep-th]}}.

\bibitem{Aharony:1997gp}
O.~Aharony, ``{IR duality in d = 3 N=2 supersymmetric USp(2N(c)) and U(N(c))
	gauge theories},''
\href{http://dx.doi.org/10.1016/S0370-2693(97)00530-3}{{\em Phys. Lett.}
	{\bfseries B404} (1997) 71--76},
\href{http://arxiv.org/abs/hep-th/9703215}{{\ttfamily arXiv:hep-th/9703215
		[hep-th]}}.

\bibitem{Aharony:2015mjs}
O.~Aharony, ``{Baryons, monopoles and dualities in Chern-Simons-matter
	theories},'' \href{http://dx.doi.org/10.1007/JHEP02(2016)093}{{\em JHEP}
	{\bfseries 02} (2016) 093},
\href{http://arxiv.org/abs/1512.00161}{{\ttfamily arXiv:1512.00161 [hep-th]}}.

\bibitem{Seiberg:2016gmd}
N.~Seiberg, T.~Senthil, C.~Wang, and E.~Witten, ``{A Duality Web in 2+1
	Dimensions and Condensed Matter Physics},''
\href{http://dx.doi.org/10.1016/j.aop.2016.08.007}{{\em Annals Phys.}
	{\bfseries 374} (2016) 395--433},
\href{http://arxiv.org/abs/1606.01989}{{\ttfamily arXiv:1606.01989 [hep-th]}}.

\bibitem{Karch:2016sxi}
A.~Karch and D.~Tong, ``{Particle-Vortex Duality from 3d Bosonization},''
\href{http://dx.doi.org/10.1103/PhysRevX.6.031043}{{\em Phys. Rev.}
	{\bfseries X6} no.~3, (2016) 031043},
\href{http://arxiv.org/abs/1606.01893}{{\ttfamily arXiv:1606.01893 [hep-th]}}.

\bibitem{Benini:2017dus}
F.~Benini, P.-S. Hsin, and N.~Seiberg, ``{Comments on global symmetries,
	anomalies, and duality in (2 + 1)d},''
\href{http://dx.doi.org/10.1007/JHEP04(2017)135}{{\em JHEP} {\bfseries 04}
	(2017) 135},
\href{http://arxiv.org/abs/1702.07035}{{\ttfamily arXiv:1702.07035
		[cond-mat.str-el]}}.

\bibitem{PhysRevD.94.085009}
S.~Kachru, M.~Mulligan, G.~Torroba, and H.~Wang, ``Bosonization and mirror
symmetry,'' \href{http://dx.doi.org/10.1103/PhysRevD.94.085009}{{\em Phys.
		Rev. D} {\bfseries 94} (Oct, 2016) 085009}.
\url{https://link.aps.org/doi/10.1103/PhysRevD.94.085009}.

\bibitem{PhysRevLett.118.011602}
S.~Kachru, M.~Mulligan, G.~Torroba, and H.~Wang, ``Nonsupersymmetric dualities
from mirror symmetry,''
\href{http://dx.doi.org/10.1103/PhysRevLett.118.011602}{{\em Phys. Rev.
		Lett.} {\bfseries 118} (Jan, 2017) 011602}.
\url{https://link.aps.org/doi/10.1103/PhysRevLett.118.011602}.

\bibitem{Kapustin:2011vz}
A.~Kapustin, H.~Kim, and J.~Park, ``{Dualities for 3d Theories with Tensor
	Matter},'' \href{http://dx.doi.org/10.1007/JHEP12(2011)087}{{\em JHEP}
	{\bfseries 12} (2011) 087},
\href{http://arxiv.org/abs/1110.2547}{{\ttfamily arXiv:1110.2547 [hep-th]}}.

\bibitem{Armoni:2017jkl}
A.~Armoni and V.~Niarchos, ``{Phases of QCD$_3$ from Non-SUSY Seiberg Duality
	and Brane Dynamics},''
\href{http://dx.doi.org/10.1103/PhysRevD.97.106001}{{\em Phys. Rev.}
	{\bfseries D97} no.~10, (2018) 106001},
\href{http://arxiv.org/abs/1711.04832}{{\ttfamily arXiv:1711.04832 [hep-th]}}.

\bibitem{Jensen:2017xbs}
K.~Jensen and A.~Karch, ``{Embedding three-dimensional bosonization dualities
	into string theory},'' \href{http://dx.doi.org/10.1007/JHEP12(2017)031}{{\em
		JHEP} {\bfseries 12} (2017) 031},
\href{http://arxiv.org/abs/1709.07872}{{\ttfamily arXiv:1709.07872 [hep-th]}}.

\bibitem{Chen:2017lkr}
J.-Y. Chen, J.~H. Son, C.~Wang, and S.~Raghu, ``{Exact Boson-Fermion Duality on
	a 3D Euclidean Lattice},''
\href{http://dx.doi.org/10.1103/PhysRevLett.120.016602}{{\em Phys. Rev.
		Lett.} {\bfseries 120} no.~1, (2018) 016602},
\href{http://arxiv.org/abs/1705.05841}{{\ttfamily arXiv:1705.05841
		[cond-mat.str-el]}}.

\bibitem{Wang:2017txt}
C.~Wang, A.~Nahum, M.~A. Metlitski, C.~Xu, and T.~Senthil, ``{Deconfined
	quantum critical points: symmetries and dualities},''
\href{http://dx.doi.org/10.1103/PhysRevX.7.031051}{{\em Phys. Rev.}
	{\bfseries X7} no.~3, (2017) 031051},
\href{http://arxiv.org/abs/1703.02426}{{\ttfamily arXiv:1703.02426
		[cond-mat.str-el]}}.

\bibitem{Komargodski:2017keh}
Z.~Komargodski and N.~Seiberg, ``{A symmetry breaking scenario for
	QCD$_{3}$},'' \href{http://dx.doi.org/10.1007/JHEP01(2018)109}{{\em JHEP}
	{\bfseries 01} (2018) 109},
\href{http://arxiv.org/abs/1706.08755}{{\ttfamily arXiv:1706.08755 [hep-th]}}.

\bibitem{Aharony:2016jvv}
O.~Aharony, F.~Benini, P.-S. Hsin, and N.~Seiberg, ``{Chern-Simons-matter
	dualities with $SO$ and $USp$ gauge groups},''
\href{http://dx.doi.org/10.1007/JHEP02(2017)072}{{\em JHEP} {\bfseries 02}
	(2017) 072},
\href{http://arxiv.org/abs/1611.07874}{{\ttfamily arXiv:1611.07874
		[cond-mat.str-el]}}.

\bibitem{PhysRevB.95.205137}
M.~A. Metlitski, A.~Vishwanath, and C.~Xu, ``Duality and bosonization of
$(2+1)$-dimensional majorana fermions,''
\href{http://dx.doi.org/10.1103/PhysRevB.95.205137}{{\em Phys. Rev. B}
	{\bfseries 95} (5, 2017) 205137}.
\url{https://link.aps.org/doi/10.1103/PhysRevB.95.205137}.

\bibitem{Aharony:2012nh}
O.~Aharony, G.~Gur-Ari, and R.~Yacoby, ``{Correlation Functions of Large N
	Chern-Simons-Matter Theories and Bosonization in Three Dimensions},''
\href{http://dx.doi.org/10.1007/JHEP12(2012)028}{{\em JHEP} {\bfseries 12}
	(2012) 028},
\href{http://arxiv.org/abs/1207.4593}{{\ttfamily arXiv:1207.4593 [hep-th]}}.

\bibitem{Aharony:2011jz}
O.~Aharony, G.~Gur-Ari, and R.~Yacoby, ``{d=3 Bosonic Vector Models Coupled to
	Chern-Simons Gauge Theories},''
\href{http://dx.doi.org/10.1007/JHEP03(2012)037}{{\em JHEP} {\bfseries 03}
	(2012) 037},
\href{http://arxiv.org/abs/1110.4382}{{\ttfamily arXiv:1110.4382 [hep-th]}}.

\bibitem{Giombi:2011kc}
S.~Giombi, S.~Minwalla, S.~Prakash, S.~P. Trivedi, S.~R. Wadia, and X.~Yin,
``{Chern-Simons Theory with Vector Fermion Matter},''
\href{http://dx.doi.org/10.1140/epjc/s10052-012-2112-0}{{\em Eur. Phys. J.}
	{\bfseries C72} (2012) 2112},
\href{http://arxiv.org/abs/1110.4386}{{\ttfamily arXiv:1110.4386 [hep-th]}}.

\bibitem{Cordova:2017vab}
C.~Cordova, P.-S. Hsin, and N.~Seiberg, ``{Global Symmetries, Counterterms, and
	Duality in Chern-Simons Matter Theories with Orthogonal Gauge Groups},''
\href{http://dx.doi.org/10.21468/SciPostPhys.4.4.021}{{\em SciPost Phys.}
	{\bfseries 4} (2018) 021},
\href{http://arxiv.org/abs/1711.10008}{{\ttfamily arXiv:1711.10008 [hep-th]}}.

\bibitem{PhysRevLett.47.1556}
C.~Dasgupta and B.~I. Halperin, ``Phase transition in a lattice model of
superconductivity,''
\href{http://dx.doi.org/10.1103/PhysRevLett.47.1556}{{\em Phys. Rev. Lett.}
	{\bfseries 47} (11, 1981) 1556--1560}.
\url{https://link.aps.org/doi/10.1103/PhysRevLett.47.1556}.

\bibitem{PESKIN1978122}
M.~E. Peskin, ``Mandelstam-'t hooft duality in abelian lattice models,''
\href{http://dx.doi.org/https://doi.org/10.1016/0003-4916(78)90252-X}{{\em
		Annals of Physics} {\bfseries 113} no.~1, (1978) 122 -- 152}.
\url{http://www.sciencedirect.com/science/article/pii/000349167890252X}.

\bibitem{2012arXiv1201.4393B}
M.~{Barkeshli} and J.~{McGreevy}, ``{A continuous transition between fractional
	quantum Hall and superfluid states},'' {\em ArXiv e-prints} (1, 2012) ,
\href{http://arxiv.org/abs/1201.4393}{{\ttfamily arXiv:1201.4393
		[cond-mat.str-el]}}.

\bibitem{Son:2015xqa}
D.~T. Son, ``{Is the Composite Fermion a Dirac Particle?},''
\href{http://dx.doi.org/10.1103/PhysRevX.5.031027}{{\em Phys. Rev.}
	{\bfseries X5} no.~3, (2015) 031027},
\href{http://arxiv.org/abs/1502.03446}{{\ttfamily arXiv:1502.03446
		[cond-mat.mes-hall]}}.

\bibitem{2015PhRvXo5d1031W}
C.~{Wang} and T.~{Senthil}, ``{Dual Dirac Liquid on the Surface of the Electron
	Topological Insulator},''
\href{http://dx.doi.org/10.1103/PhysRevX.5.041031}{{\em Physical Review X}
	{\bfseries 5} no.~4, (10, 2015) 041031},
\href{http://arxiv.org/abs/1505.05141}{{\ttfamily arXiv:1505.05141
		[cond-mat.str-el]}}.

\bibitem{2016PhRvXv6c1026P}
A.~C. {Potter}, M.~{Serbyn}, and A.~{Vishwanath}, ``{Thermoelectric Transport
	Signatures of Dirac Composite Fermions in the Half-Filled Landau Level},''
\href{http://dx.doi.org/10.1103/PhysRevX.6.031026}{{\em Physical Review X}
	{\bfseries 6} no.~3, (7, 2016) 031026},
\href{http://arxiv.org/abs/1512.06852}{{\ttfamily arXiv:1512.06852
		[cond-mat.str-el]}}.

\bibitem{2016PhRvBa94x5107W}
C.~{Wang} and T.~{Senthil}, ``{Composite Fermi liquids in the lowest Landau
	level},'' \href{http://arxiv.org/abs/1604.06807}{{\ttfamily arXiv:1604.06807
		[cond-mat.str-el]}}.

\bibitem{Murugan:2016zal}
J.~Murugan and H.~Nastase, ``{Particle-vortex duality in topological insulators
	and superconductors},'' \href{http://dx.doi.org/10.1007/JHEP05(2017)159}{{\em
		JHEP} {\bfseries 05} (2017) 159},
\href{http://arxiv.org/abs/1606.01912}{{\ttfamily arXiv:1606.01912 [hep-th]}}.

\bibitem{Jensen:2017bjo}
K.~Jensen, ``{A master bosonization duality},''
\href{http://dx.doi.org/10.1007/JHEP01(2018)031}{{\em JHEP} {\bfseries 01}
	(2018) 031},
\href{http://arxiv.org/abs/1712.04933}{{\ttfamily arXiv:1712.04933 [hep-th]}}.

\bibitem{Gur-Ari:2015pca}
G.~Gur-Ari and R.~Yacoby, ``{Three Dimensional Bosonization From
	Supersymmetry},'' \href{http://dx.doi.org/10.1007/JHEP11(2015)013}{{\em JHEP}
	{\bfseries 11} (2015) 013},
\href{http://arxiv.org/abs/1507.04378}{{\ttfamily arXiv:1507.04378 [hep-th]}}.

\bibitem{GomisUnpublished}
D.~Delmastro and J.~Gomis unpublished.

\bibitem{Nielsen:1978mp}
N.~K. Nielsen, ``{Ghost Counting in Supergravity},''
\href{http://dx.doi.org/10.1016/0550-3213(78)90009-3}{{\em Nucl. Phys.}
	{\bfseries B140} (1978) 499--509}.

\bibitem{Kallosh:1978de}
R.~E. Kallosh, ``{Modified Feynman Rules in Supergravity},''
\href{http://dx.doi.org/10.1016/0550-3213(78)90340-1}{{\em Nucl. Phys.}
	{\bfseries B141} (1978) 141--152}.

\bibitem{Witten:1999ds}
E.~Witten, ``{Supersymmetric index of three-dimensional gauge theory},''
\href{http://arxiv.org/abs/hep-th/9903005}{{\ttfamily arXiv:hep-th/9903005
		[hep-th]}}.

\bibitem{naculich1990group}
S.~G. Naculich, H.~Riggs, and H.~Schnitzer, ``Group-level duality in wzw models
and chern-simons theory,'' {\em Physics Letters B} {\bfseries 246} no.~3-4,
(1990) 417--422.

\bibitem{mlawer1991group}
E.~J. Mlawer, S.~G. Naculich, H.~A. Riggs, and H.~J. Schnitzer, ``Group-level
duality of wzw fusion coefficients and chern-simons link observables,'' {\em
	Nuclear Physics B} {\bfseries 352} no.~3, (1991) 863--896.

\bibitem{Hsin:2016blu}
P.-S. Hsin and N.~Seiberg, ``{Level/rank Duality and Chern-Simons-Matter
	Theories},'' \href{http://dx.doi.org/10.1007/JHEP09(2016)095}{{\em JHEP}
	{\bfseries 09} (2016) 095},
\href{http://arxiv.org/abs/1607.07457}{{\ttfamily arXiv:1607.07457 [hep-th]}}.

\bibitem{Armoni:2005sp}
A.~Armoni and T.~J. Hollowood, ``{Sitting on the domain walls of N=1 super
	Yang-Mills},'' \href{http://dx.doi.org/10.1088/1126-6708/2005/07/043}{{\em
		JHEP} {\bfseries 07} (2005) 043},
\href{http://arxiv.org/abs/hep-th/0505213}{{\ttfamily arXiv:hep-th/0505213
		[hep-th]}}.

\bibitem{Armoni:2006ee}
A.~Armoni and T.~J. Hollowood, ``{Interactions of domain walls of SUSY
	Yang-Mills as D-branes},''
\href{http://dx.doi.org/10.1088/1126-6708/2006/02/072}{{\em JHEP} {\bfseries
		02} (2006) 072},
\href{http://arxiv.org/abs/hep-th/0601150}{{\ttfamily arXiv:hep-th/0601150
		[hep-th]}}.

\bibitem{PhysRevD.10.1246}
H.~Georgi and A.~Pais, ``$\mathrm{CP}$ violation as a quantum effect,''
\href{http://dx.doi.org/10.1103/PhysRevD.10.1246}{{\em Phys. Rev. D}
	{\bfseries 10} (8, 1974) 1246--1250}.
\url{https://link.aps.org/doi/10.1103/PhysRevD.10.1246}.

\bibitem{Dunne:1998qy}
G.~V. Dunne, ``{Aspects of Chern-Simons theory},'' in {\em {Topological Aspects
		of Low-dimensional Systems: Proceedings, Les Houches Summer School of
		Theoretical Physics, Session 69: Les Houches, France, July 7-31 1998}}.
\newblock 1998.
\newblock
\href{http://arxiv.org/abs/hep-th/9902115}{{\ttfamily arXiv:hep-th/9902115
		[hep-th]}}.
\newblock

\bibitem{benini2018n}
F.~Benini and S.~Benvenuti, ``N= 1 qed in 2+ 1 dimensions: Dualities and
enhanced symmetries,'' {\em arXiv preprint arXiv:1804.05707} (2018) .

\bibitem{Aharony:2014uya}
O.~Aharony and D.~Fleischer, ``{IR Dualities in General 3d Supersymmetric SU(N)
	QCD Theories},'' \href{http://dx.doi.org/10.1007/JHEP02(2015)162}{{\em JHEP}
	{\bfseries 02} (2015) 162},
\href{http://arxiv.org/abs/1411.5475}{{\ttfamily arXiv:1411.5475 [hep-th]}}.

\bibitem{Kapustin:2009kz}
A.~Kapustin, B.~Willett, and I.~Yaakov, ``{Exact Results for Wilson Loops in
	Superconformal Chern-Simons Theories with Matter},''
\href{http://dx.doi.org/10.1007/JHEP03(2010)089}{{\em JHEP} {\bfseries 03}
	(2010) 089},
\href{http://arxiv.org/abs/0909.4559}{{\ttfamily arXiv:0909.4559 [hep-th]}}.

\bibitem{Hama:2010av}
N.~Hama, K.~Hosomichi, and S.~Lee, ``{Notes on SUSY Gauge Theories on
	Three-Sphere},'' \href{http://dx.doi.org/10.1007/JHEP03(2011)127}{{\em JHEP}
	{\bfseries 03} (2011) 127},
\href{http://arxiv.org/abs/1012.3512}{{\ttfamily arXiv:1012.3512 [hep-th]}}.

\bibitem{Jafferis:2010un}
D.~L. Jafferis, ``{The Exact Superconformal R-Symmetry Extremizes Z},''
\href{http://dx.doi.org/10.1007/JHEP05(2012)159}{{\em JHEP} {\bfseries 05}
	(2012) 159},
\href{http://arxiv.org/abs/1012.3210}{{\ttfamily arXiv:1012.3210 [hep-th]}}.

\bibitem{Gates:1983nr}
S.~J. Gates, M.~T. Grisaru, M.~Rocek, and W.~Siegel, ``{Superspace Or One
	Thousand and One Lessons in Supersymmetry},'' {\em Front. Phys.} {\bfseries
	58} (1983) 1--548,
\href{http://arxiv.org/abs/hep-th/0108200}{{\ttfamily arXiv:hep-th/0108200
		[hep-th]}}.

\bibitem{karlhede1987supersymmetric}
A.~Karlhede, U.~Lindstrom, M.~Rocek, and P.~van Nieuwenhuizen, ``Supersymmetric
vector-vector duality,'' {\em Classical and Quantum Gravity} {\bfseries 4}
no.~3, (1987) 549.

\bibitem{hori2003mirror}
K.~Hori, {\em Mirror symmetry}, vol.~1.
\newblock American Mathematical Soc., 2003.	
	
\end{thebibliography}
\providecommand{\href}[2]{#2}\begingroup\raggedright

\endgroup
\end{document}